\documentclass[11pt]{article}

\usepackage[nottoc,notlot,notlof]{tocbibind}

\usepackage[T1]{fontenc}
 \usepackage[english]{babel}
\usepackage{layout}
\usepackage{natbib}
\usepackage{amsmath}
\usepackage{mathrsfs}
\usepackage{amssymb}
\usepackage{changes}
\usepackage{graphicx}
\usepackage[euler]{textgreek}
\usepackage{ucs}
\usepackage{float}
\usepackage{lipsum}
\usepackage{listing}
\usepackage{bibentry}
\usepackage{appendix}
\usepackage{upgreek}
\usepackage{listings}
\usepackage[makeroom]{cancel}
\usepackage{setspace}
\usepackage{enumitem}
\usepackage{cite}
\usepackage{textcomp}
\usepackage{lineno}
\usepackage[colorlinks=true,citecolor=blue]{hyperref}%
\usepackage[a4paper,top=2cm, bottom=2cm, left=2cm, right=2cm]{geometry}
\usepackage{todonotes}
\usepackage{siunitx}
\usepackage{caption}
\usepackage[ruled,vlined]{algorithm2e}
\usepackage{relsize} 

\usepackage{ragged2e}
\usepackage[scaled=.7]{beramono}
\usepackage{pdfpages}
\usepackage{authblk}
\usepackage{cancel}

\definechangesauthor[color=cyan]{JJ}
\definechangesauthor[color=brown]{ND}
\definechangesauthor[color=red]{WS}
\definechangesauthor[color=pink]{MP}

\newcommand{\x}{\boldsymbol{x}}
\newcommand{\X}{\boldsymbol{X}}
\newcommand{\z}{\boldsymbol{z}}
\newcommand{\Z}{\boldsymbol{Z}}
\newcommand{\y}{\boldsymbol{y}}

\newcommand{\disc}{\boldsymbol{d}}
\newcommand{\D}{\boldsymbol{D}}
\newcommand{\PAD}{PAD}
\newcommand{\nPAD}{{\xcancel{P}}AD}
\newcommand{\PnAD}{P{\xcancel{A}}D}

\makeatletter

\title{Learning cortical representations through perturbed and adversarial dreaming}
\author[1\footnote{Correspondence: nicolas.deperrois@unibe.ch}]{Nicolas Deperrois}
\author[1,2]{Mihai A.~Petrovici}
\author[1\footnote{Joint senior authorship.}]{Walter Senn}
\author[1$^\dagger$]{Jakob Jordan}

\affil[1]{Department of Physiology, University of Bern}
\affil[2]{Kirchhoff-Institute for Physics, Heidelberg University}

\date{}

\begin{document}

\maketitle

\begin{abstract}
Humans and other animals learn to extract general concepts from sensory experience without extensive teaching.
This ability is thought to be facilitated by offline states like sleep where previous experiences are systemically replayed.
However, the characteristic creative nature of dreams suggests that learning semantic representations may go beyond merely replaying previous experiences.
We support this hypothesis by implementing a cortical architecture inspired by generative adversarial networks (GANs).
Learning in our model is organized across three different global brain states mimicking wakefulness, NREM and REM sleep, optimizing different, but complementary objective functions.
We train the model on standard datasets of natural images and evaluate the quality of the learned representations.
Our results suggest that generating new, virtual sensory inputs via adversarial dreaming during REM sleep is essential for extracting semantic concepts, while replaying episodic memories via perturbed dreaming during NREM sleep improves the robustness of latent representations.
The model provides a new computational perspective on sleep states, memory replay and dreams and suggests a cortical implementation of GANs.
\end{abstract}


\section{Introduction}

After just a single night of bad sleep, we are acutely aware of the importance of sleep for orderly body and brain function.
In fact, it has become clear that sleep serves multiple crucial physiological functions \citep{siegel_sleep_2009, xie2013sleep}, and growing evidence highlights its impact on cognitive processes \citep{walker_role_2009}.
Yet, a lot remains unknown about the precise contribution of sleep, and in particular dreams, on normal brain function.

One remarkable cognitive ability of humans and other animals lies in the extraction of general concepts and statistical regularities from sensory experience without extensive teaching \citep{bergelson_at_2012}.
Such regularities in the sensorium are reflected on the neuronal level in invariant object-specific representations in high-level areas of the visual cortex \citep{grill-spector_2001_the, hung_fast_2005, dicarlo_how_2012} on which downstreams areas can operate.
These so called semantic representations are progressively constructed and enriched over an organism's lifetime \citep{tenenbaum_how_2011, yee_semantic_2013} and their emergence is hypothesized to be facilitated by offline states such as sleep \citep{dudai_consolidation_2015}.

Previously, several cortical models have been proposed to explain how offline states could contribute to the emergence of high-level, semantic representations. 
Stochastic hierarchical models which learn to maximize the likelihood of observed data under a generative model such as the Helmholtz machine \citep{dayan_helmholtz_1995} and the closely related Wake-Sleep algorithm \citep{hinton_wake-sleep_1995, bornschein2015reweighted} have demonstrated the potential of combining online and offline states to learn semantic representations.
However, these models do not leverage offline states to improve their generative model but are explicitly trained to reproduce sensory inputs during wakefulness.
In contrast, most dreams during REM sleep exhibit realistic imagery beyond past sensory experience \citep{fosse_dreaming_2003, nir_dreaming_2010, wamsley_dreaming_2014} suggesting learning principles which go beyond mere reconstructions.

In parallel, cognitive models inspired by psychological studies of sleep proposed a "trace transformation theory" where semantic knowledge is actively extracted in the cortex from replayed hippocampal episodic memories \citep{nadel_memory_1997, winocur_memory_2010, lewis_overlapping_2011}.
However, these models lack a mechanistic implementation compatible with cortical structures and only consider the replay of waking activity during sleep.

Recently, implicit generative models which do not explicitly try to reconstruct observed sensory inputs, and in particular generative adversarial networks \citep[GANs;][]{goodfellow2014generative}, have been successfully applied in machine learning to generate new but realistic data from random patterns.
This ability has been shown to be accompanied by the learning of disentangled and semantically meaningful representations \citep{radford_unsupervised_2015, donahue_adversarial_2016, liu_self-supervised_2021}.
They thus may provide computational principles for learning cortical semantic representations during offline states by generating previously unobserved sensory content as reported from dream experiences.

Most dreams experienced during rapid-eye-movement (REM) sleep only incorporate fragments of previous waking experience, often intermingled with past memories \citep{schwartz_are_2003}. 
Suprisingly, such random combinations of memory fragments often results in visual experiences which are perceived as highly structured and realistic by the dreamer. 
The striking similarity between the inner world of dreams and the external world of wakefulness suggests that the brain actively creates novel experiences by rearranging stored episodic patterns in a meaningful manner  \citep{nir_dreaming_2010}.
A few hypothetical functions were attributed to this phenomenon, such as enhancing creative problem solving by building novel associations between unrelated memory elements \citep{cai_rem_2009, llewellyn_crossing_2016, lewis_how_2018}, forming internal prospective codes oriented toward future waking experiences \citep{llewellyn_dream_2016}, or refining a generative model by minimizing its complexity and improving generalization \citep{hobson_virtual_2014, hoel_overfitted_2021}.
However, these theories do not consider the role of dreams for a more basic function, such as the formation of semantic cortical representations. 

Here, we propose that dreams, and in particular their creative combination of episodic memories, play an essential role in forming semantic representations over the course of development.
The formation of representations which abstract away redundant information from sensory input and which can thus be easily used by downstream areas is an important basis for memory semantization.
To support this hypothesis, we introduce a new, functional model of cortical representation learning.
The central ingredient of our model is a creative generative process via feedback from higher to lower cortical areas which mimics dreaming during REM sleep. 
This generative process is trained to produce more realistic virtual sensory experience in an adversarial fashion by trying to fool an internal mechanism distinguishing low-level activities between wakefulness and REM sleep.
Intuitively, generating new but realistic sensory experiences, instead of merely reconstructing previous observations, requires the brain to understand the composition of its sensorium.
In line with transformation theories, this suggests that cortical representations should carry semantic, decontextualized gist information.

\begin{figure}[t]
  {\centering
    \includegraphics[width=0.9\textwidth]{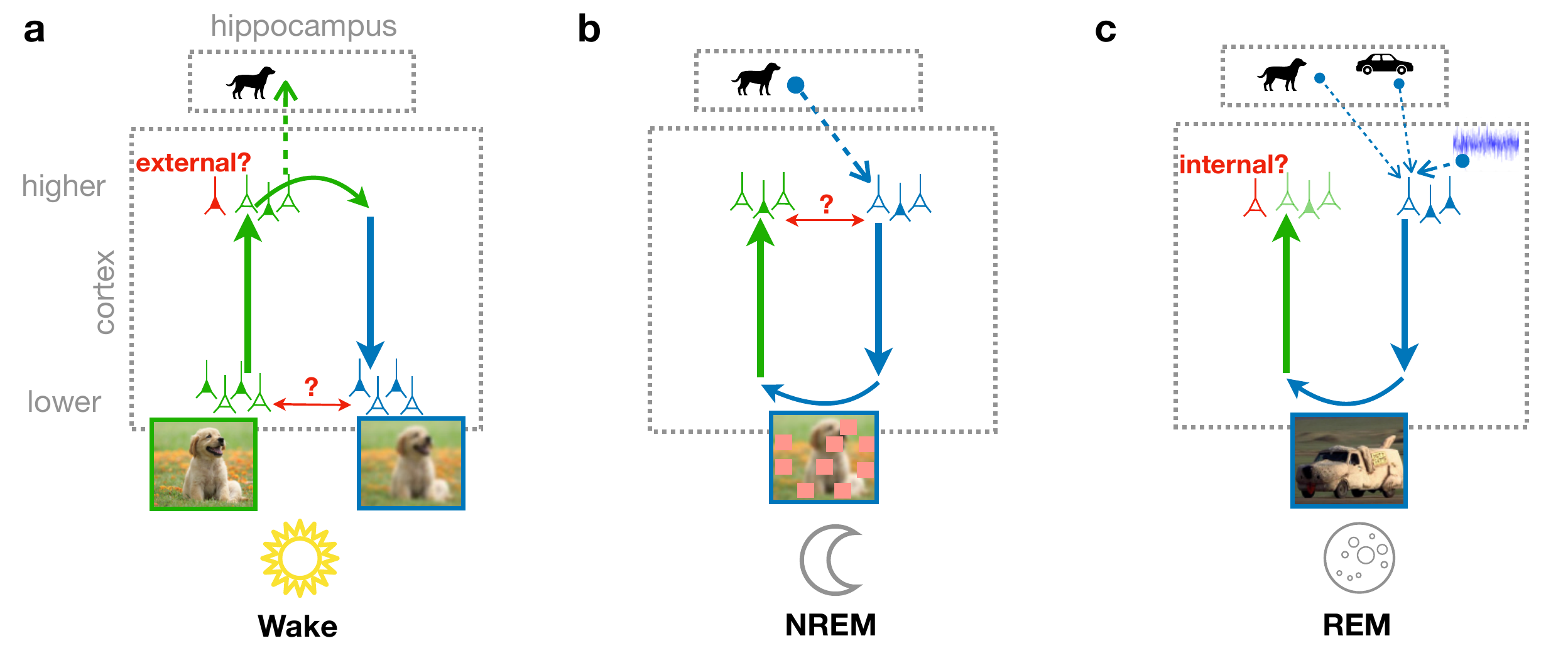}
    \caption{
      \small
      {\bf Cortical representation learning through perturbed and adversarial dreaming (\PAD).}
      {\bf (a)} During wakefulness (Wake), cortical feedforward pathways learn to recognize that low-level activity is externally-driven and feedback pathways learn to reconstruct it from high-level neuronal representations. These high-level representations are stored in the hippocampus.
      {\bf (b)} During NREM sleep (NREM), feedforward pathways learn to reconstruct high-level activity patterns replayed from the hippocampus affected by low-level perturbations, referred to as perturbed dreaming.
      {\bf (c)} During REM sleep (REM), feedforward and feedback pathways operate in an adversarial fashion, referred to as adversarial dreaming.
      Feedback pathways generate virtual low-level activity from combinations of multiple hippocampal memories and spontaneous cortical activity. 
      While feedforward pathways learn to recognize low-level activity patterns as internally generated, feedback pathways learn to fool feedforward pathways. 
      \label{fig:intro-three-phases}
      }
  }
\end{figure}

We implement this model in a cortical architecture with hierarchically organized forward and backward pathways, loosely inspired by GANs.
The connectivity of the model is adapted by gradient-based synaptic plasticity, optimizing different, but complementary objective functions depending on the brain's global state.
During wakefulness, the model learns to recognize that low-level activity is externally-driven, stores high-level representations in the hippocampus, and tries to predict low-level from high-level activity (Fig.~\ref{fig:intro-three-phases}a).
During NREM sleep, the model learns to reconstruct replayed high-level activity patterns from generated low-level activity, perturbed by virtual occlusions, referred to as perturbed dreaming (Fig.~\ref{fig:intro-three-phases}b).
During REM sleep, the model learns to generate realistic low-level activity patterns from random combinations of several hippocampal memories and spontaneous cortical activity, while simultaneously learning to distinguish these virtual experiences from externally-driven waking experiences, referred to as adversarial dreaming (Fig.~\ref{fig:intro-three-phases}c).
Together with the wakefulness, the two sleep states, NREM and REM, jointly implement our model of Perturbed and Adversarial Dreaming (\PAD).

Over the course of learning, constrained by its architecture and the prior distribution of latent activities, our cortical model trained on natural images develops rich latent representations along with the capacity to generate plausible early sensory activities. 
We demonstrate that adversarial dreaming during REM sleep is essential for learning representations organized according to object semantics, which are improved and robustified by perturbed dreaming during NREM sleep.
Together, our results demonstrate a potential role of dreams and suggest complementary functions of REM and NREM sleep in cortical representation learning.

\section{Results}

\subsection{Complementary objectives for wakefulness, NREM and REM sleep}

\begin{figure}[t!]
    {\centering
    \includegraphics[width=1.0\textwidth]{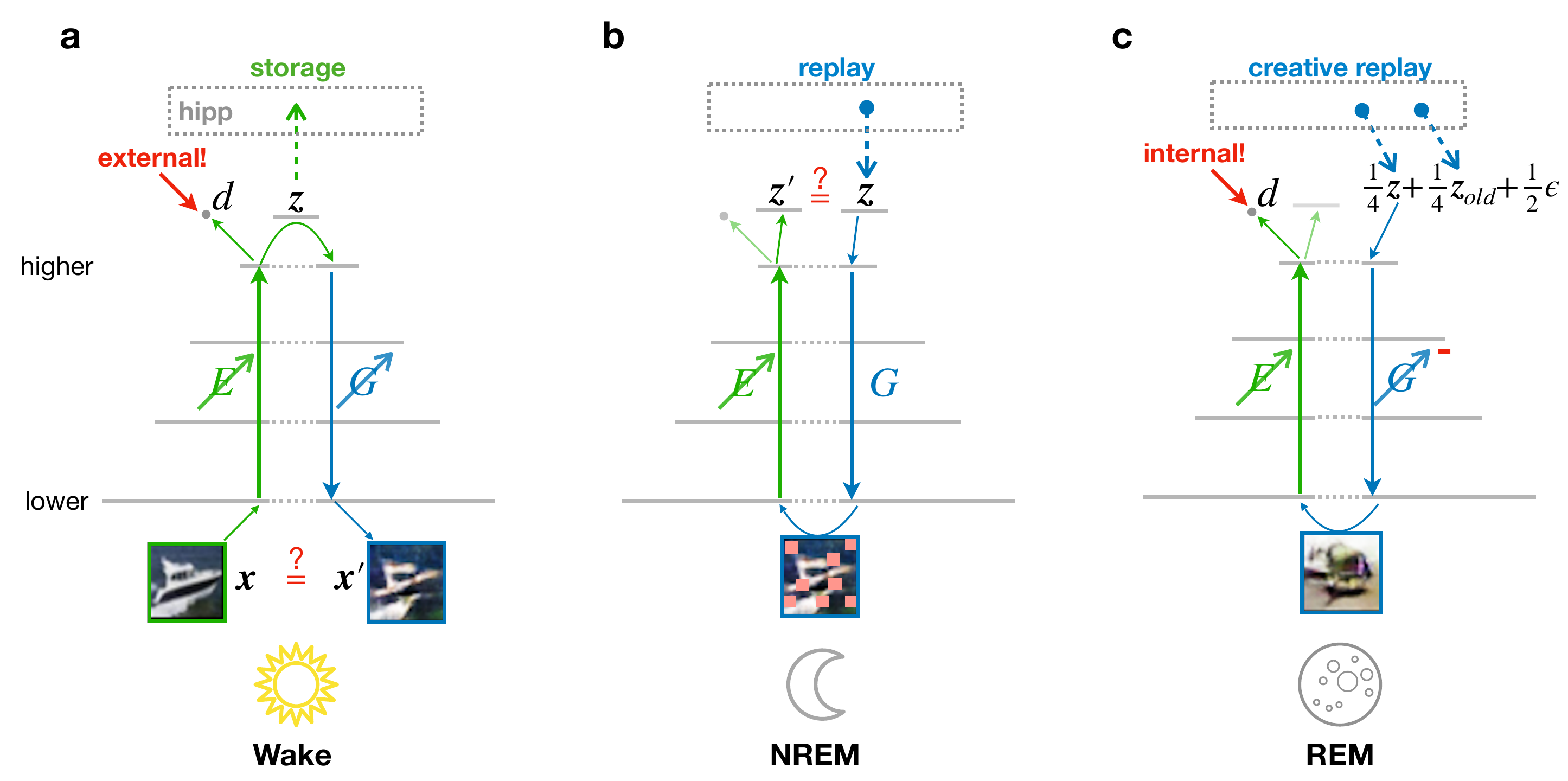}
    \caption{{\bf Different objectives during wakefulness, NREM, and REM sleep govern the organization of feedforward and feedback pathways in \PAD}
    The variable $\x$ corresponds to 32x32 image, $\z$ is a $256$-dimensional vector representing the latent layer (higher sensory cortex).
    Encoder ($E$, green) and generator ($G$, blue) networks project bottom-up and top-down signals between lower and higher sensory areas.
    An oblique arrow ($\nearrow$) indicates that learning occurs in a given pathway.
    \textbf{(a)} During Wake, low-level activities $\x$ are reconstructed.
    At the same time, $E$ learns to classify low-level activity as external (red target `external!') with its output discriminator $d$.
    The obtained latent representations $\z$ are stored in the hippocampus.
    {\bf (b)} During NREM, the activity $\z$ stored during wakefulness is replayed from the hippocampal memory and regenerates visual input from the previous day perturbed by occlusions, modelled by squares of various sizes applied along the generated low-level activity with a certain probability (see Methods).
    In this phase, $E$ adapts to reproduce the replayed latent activity.
    \textbf{(c)} During REM, convex combinations of multiple random hippocampal memories ($\z$ and $\z_{\text{old}}$) and spontaneous cortical activity ($\epsilon$), here with specific prefactors, generate a virtual activity in lower areas.  
    While the encoder learns to classify this activity as internal (red target `internal!'), the generator adversarially learns to generate visual inputs that would be classified as external.
    The red minus on $G$ indicates the inverted plasticity implementing this adversarial training.
    }\label{fig:sketch}
  }
\end{figure}

We consider an abstract model of the visual ventral pathway consisting of multiple, hierarchically organized cortical areas, with a feedforward pathway, or encoder, transforming neuronal activities from lower to higher areas (Fig.~\ref{fig:sketch}, $E$). 
These high-level activities are compressed representations of low-level activities and are called latent representations, here denoted by $\z$.
In addition to this feedforward pathway, we similarly model a feedback pathway, or generator, projecting from higher to lower areas (Fig.~\ref{fig:sketch}, $G$).
These two pathways are supported by a simple hippocampal module which can store and replay latent representations.
Three different global brain states are considered: wakefulness (Wake), non-REM sleep (NREM) and REM sleep (REM).
We focus on the functional role of these phases while abstracting away dynamical features such as bursts, spindles or slow waves \citep{leger_slow-wave_2018}, in line with previous approaches based on goal-driven modeling which successfully predict physiological features along the ventral stream \citep{yamins_performance-optimized_2014, zhuang_unsupervised_2021}.

In our model, the three brain states only differ in their objective function and the presence or absence of external input.
Synaptic plasticity performs stochastic gradient descent on state-specific objective functions via error backpropagation \citep{lecun_deep_2015}.
We assume that efficient credit assignment is realized in the cortex, and focus on the functional consequences of our specific architecture. For potential implementations of biophysically plausible backpropagation in cortical circuits, we refer to previous work \citep[e.g.,][]{whittington_theories_2019, lillicrap_backpropagation_2020}.

During Wake (Fig.~\ref{fig:sketch}a), sensory inputs evoke activities $\x$ in lower sensory cortex which are transformed via the feedforward pathway $E$ into latent representations $\z$ in higher sensory cortex.
The hippocampal module stores these latent representations, mimicking the formation of episodic memories.
Simultaneously, the feedback pathway $G$ generates low-level activities $\x'$ from these representations.
Synaptic plasticity adapts the encoding and generative pathways ($E$ and $G$) to minimize the mismatch between externally-driven and internally-generated activities (Fig.~\ref{fig:sketch}a).
Thus, the network learns to reproduce low-level activity from abstract high-level representations.
Simultaneously, $E$ also acts as a `discriminator' with output $d$ that is trained to become active, reflecting that the low-level activity was driven by an external stimuli.
The discriminator learning during Wake is essential to drive adversarial learning during REM. 
Note that computationally the classification of low-level cortical activities into ``externally driven'' and ``internally generated'' is not different from classification into, for example, different object categories, even though conceptually they serve different purposes.
The dual use of $E$ reflects a view of cortical information processing in which several network functions are preferentially shared among a single network mimicking the ventral visual stream \citep{dicarlo_how_2012}.
This approach has been previously successfully employed in machine learning models \citep{huang_introVAE_2018, brock_neural_2017, ulyanov_it_2017, munjal_implicit_2019, bang_discriminator_2020}.

For the subsequent sleep phases, the system is disconnected from the external environment, and activity in lower sensory cortex is driven by top-down signals originating from higher areas, as previously suggested \citep{nir_dreaming_2010,aru_apical_2020}.
During NREM (Fig.~\ref{fig:sketch}b), latent representations $\z$ are recalled from the hippocampal module, corresponding to the replay of episodic memories.
These representations generate low-level activities which are perturbed by suppressing early sensory neurons, modeling the observed differences between replayed and waking activities \citep{ji_coordinated_2007}.
The encoder reconstructs latent representations from these activity patterns, and synaptic plasticity adjusts the feedforward pathway to make the latent representation of the perturbed generated activity similar to the original episodic memory.
This process defines perturbed dreaming.

During REM (Fig.~\ref{fig:sketch}c), sleep is characterized by creative dreams generating realistic virtual sensory experiences out of the combination of episodic memories \citep{fosse_dreaming_2003,lewis_how_2018}.
In \PAD, multiple random episodic memories from the hippocampal module are linearly combined and projected to cortex.
Reflecting the decreased coupling \citep{wierzynski_2009, lewis_how_2018} between hippocampus and cortex during REM sleep, these mixed representations are diluted with spontaneous cortical activity, here abstracted as Gaussian noise with zero mean and unit variance.
From this new high-level cortical representation, activity in lower sensory cortex is generated and finally passed through the feedforward pathway.
Synaptic plasticity adjusts feedforward connections $E$ to silence the activity of the discriminator output as it should learn to distinguish it from externally-evoked sensory activity.
Simultaneously, feedback connections are adjusted adversarially to generate activity patterns which appear externally-driven and thereby trick the discriminator into believing that the low-level activity was externally-driven.
This is achieved by inverting the sign of the errors that determine synaptic weight changes in the generative network.
This process defines adversarial dreaming.

The functional differences between our proposed NREM and REM sleep phases are motivated by experimental data describing a reactivation of hippocampal memories during NREM sleep and the occurrence of creative dreams during REM sleep.
In particular,  hippocampal replay has been reported during NREM sleep within sharp-wave-ripples \citep{oneill_play_2010}, also observed in the visual cortex \citep{ji_coordinated_2007}, which resembles activity from wakefulness.
Our REM sleep phase is built upon cognitive theories of REM dreams \citep{llewellyn_dream_2016, lewis_how_2018} postulating that they emerge from random combinations between episodic memory elements, sometimes remote from each other, which appear realistic for the dreamer.
This random coactivation could be caused by theta oscillations in the hippocampus during REM sleep \citep{buzsaki_theta_2002}.
The addition of cortical noise is motivated by experimental work showing reduced correlations between hippocampal and cortical activity during REM sleep \citep{wierzynski_2009}, and the occurence of ponto-geniculo-occipital (PGO) waves \citep{nelson_rem_1983} in the visual cortex often associated with generation of novel visual imagery in dreams \citep{hobson_dreaming_2000, hobson_virtual_2014}.
Furthermore, the cortical contribution in REM dreaming is supported by experimental evidence that dreaming still occurs with hippocampal damage, while reported to be less episodic-like in nature \citep{spano_dreaming_2020}.

Within our suggested framework, `dreams' arise as early sensory activity that is internally-generated via feedback pathways during offline states, and subsequently processed by feedforward pathways.
In particular, this implies that besides REM dreams, NREM dreams exist.
However, in contrast to REM dreams, which are significantly different from waking experiences \citep{fosse_dreaming_2003}, our model implies that NREM dreams are more similar to waking experiences since they are driven by single episodic memories, in contrast to REM dreams which are generated from a mixture of episodic memories.
Furthermore, the implementation of adversarial dreaming requires an internal representation of whether early sensory activity is externally or internally generated, i.e., a distinction whether a sensory experience is real or imagined.

\subsection{Dreams become more realistic over the course of learning}

\begin{figure}[t]
  {\centering
    \includegraphics[width=1.0\textwidth]{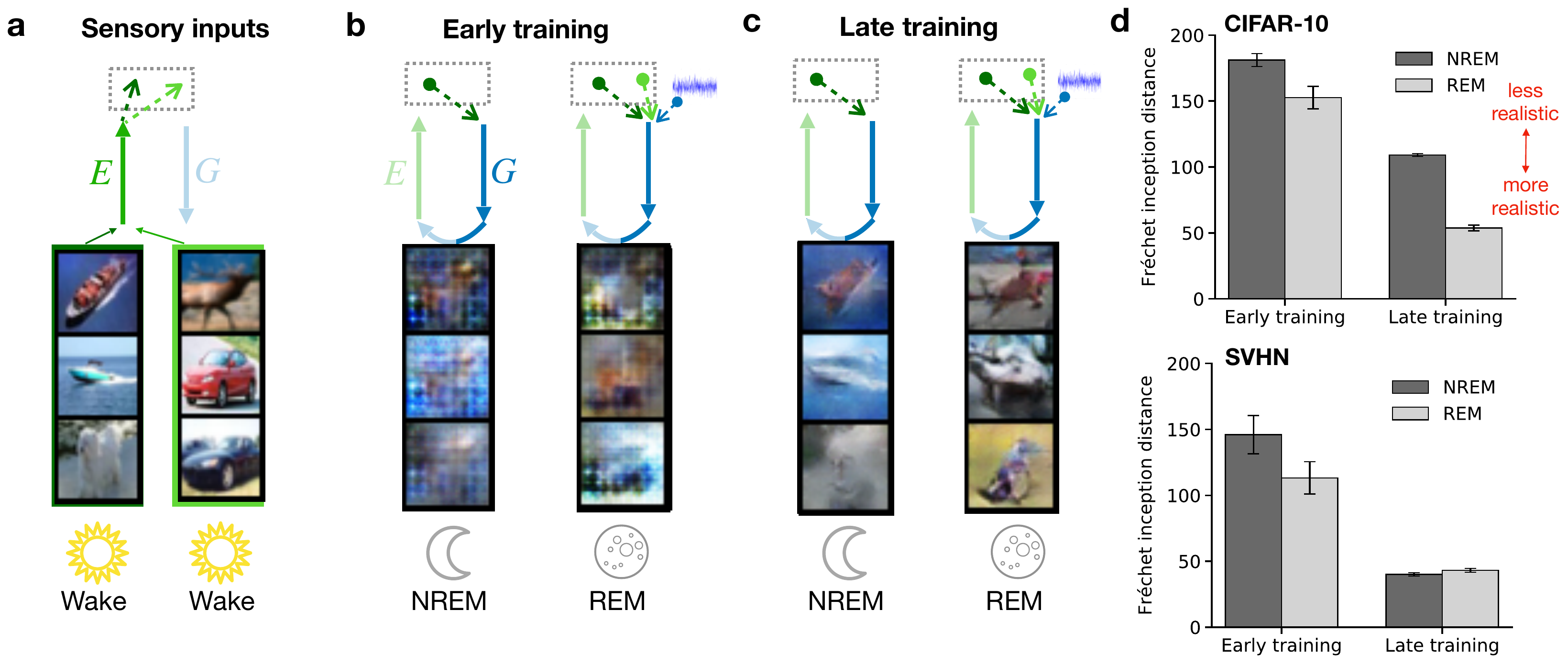}
    \caption{{\bf Both NREM and REM dreams become more realistic over the course of learning.}
      {\bf (a)} Examples of sensory inputs observed during wakefulness. Their corresponding latent representations are stored in the hippocampus.
      {\bf (b, c)} Single episodic memories (latent representations of stimuli) during NREM from the previous day and combinations of episodic memories from the two previous days during REM are recalled from hippocampus and generate early sensory activity via feedback pathways.
      This activity is shown for early (epoch 1) and late (epoch 50) training stages of the model.
      {\bf (d)} Discrepancy between externally-driven and internally-generated early sensory activity as measured by the Fr\'echet inception distance (FID) \citep{heusel_gans_2018} during NREM and REM for networks trained on CIFAR-10 (top) and SVHN (bottom).
      Lower distance reflects higher similarity between sensory-evoked and generated activity.
      Error bars indicate $\pm 1$ SEM over 4 different initial conditions.
    }\label{fig:results_dreams}
  }
\end{figure}

Dreams in our model arise from both NREM (perturbed dreaming) and REM (adversarial dreaming) phases.
In both cases, they are characterized by activity in early sensory areas generated via feedback pathways.
To illustrate learning in \PAD, we consider these low-level activities during NREM and during REM for a model with little learning experience ("early training") and a model which has experienced many wake-sleep cycles ("late training"; Fig.~\ref{fig:results_dreams}).
A single wake-sleep cycle consists of Wake, NREM and REM phases.
As an example, we train our model on a dataset of natural images \citep[CIFAR-10;][]{cifar10} and a dataset of images of house numbers \citep[SVHN;][]{svhn}.
Initially, internally-generated low-level activities during sleep do not share significant similarities with sensory-evoked activities from Wake (Fig.~\ref{fig:results_dreams}a); for example, no obvious object shapes are represented (Fig.~\ref{fig:results_dreams}b).
After plasticity has organized network connectivity over many wake-sleep cycles (50 training epochs), low-level internally-generated activity patterns resemble sensory-evoked activity (Fig.~\ref{fig:results_dreams}c).
NREM-generated activities reflect the sensory content of the episodic memory (sensory input from the previous day).
REM-generated activities are different from the sensory activities corresponding to the original episodic memories underlying them as they recombine features of sensory activities from the two previous days, but still exhibit a realistic structure.
This increase in similarity between externally-driven and internally-generated low-level activity patterns is also reflected in a decreasing Fr\'echet inception distance (Fig.~\ref{fig:results_dreams}d), a metric used to quantify the realism of generated images \citep{heusel_gans_2018}.
The increase of dreams realism, here mostly driven by a combination of reconstruction learning (Wake) and adversarial learning (Wake and REM), correlates with the development of dreams in children, that are initially plain and fail to represent objects, people, but become more realistic and structured over time \citep{foulkes_children_1999, nir_dreaming_2010}.

The \PAD~training paradigm hence leads to internally-generated low-level activity patterns that become more difficult to discern from externally-driven activities, whether they originate from single episodic memories during NREM or from noisy random combinations thereof during REM.
We will next demonstrate that the same learning process leads to the emergence of robust semantic representations.

\subsection{Adversarial dreaming during REM facilitates the emergence of semantic representations}

\begin{figure}[t]
    {\centering
    \includegraphics[width=0.9\textwidth]{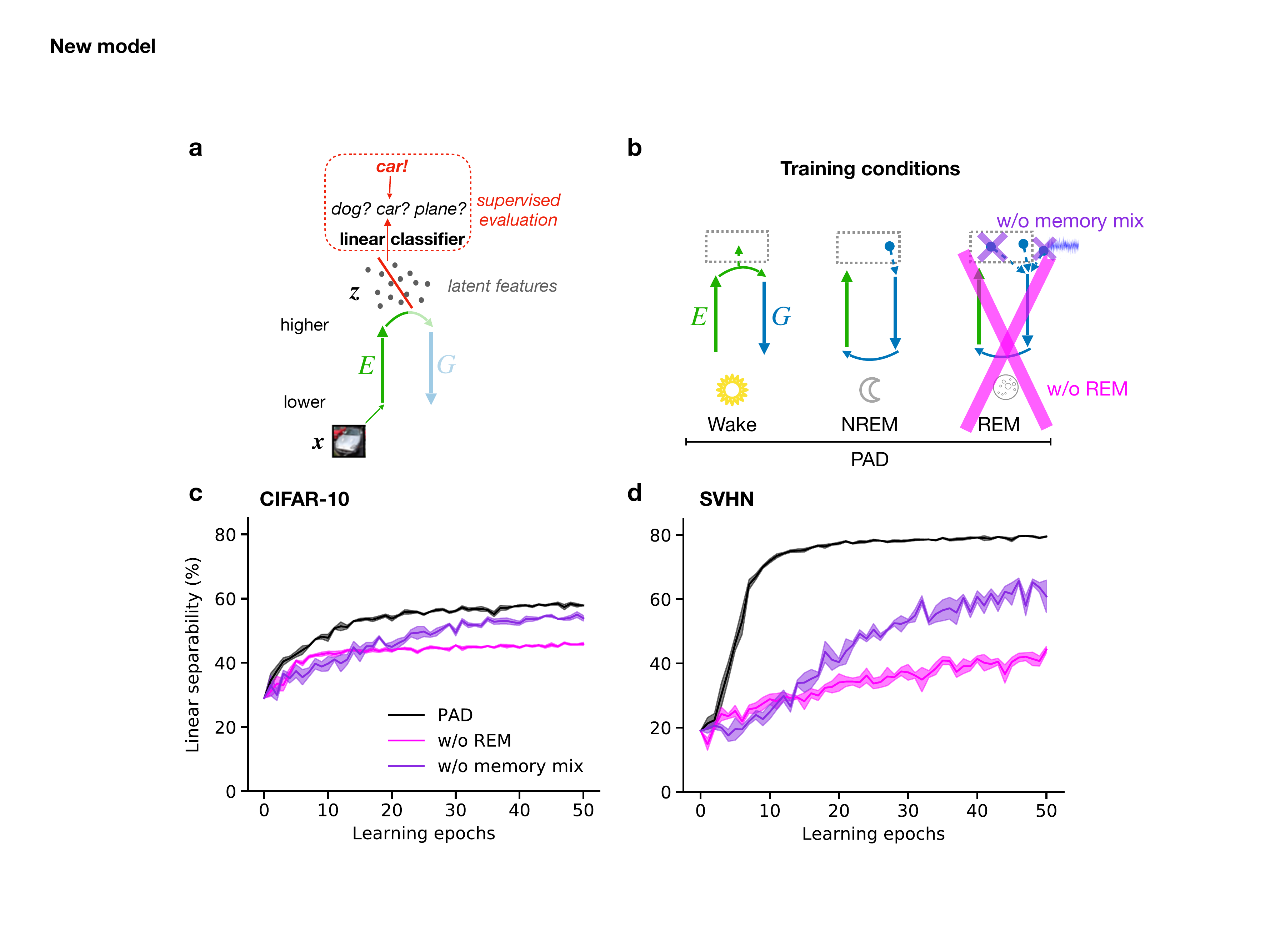}
    \caption{{\bf Adversarial dreaming during REM improves the linear separability of the latent representation.}
    \textbf{(a)} A linear classifier is trained on the latent representations $\boldsymbol{z}$ inferred from an external input $\x$ to predict its associated label (here, the category `car').
    \textbf{(b)} Training phases and pathological conditions: full model (\PAD $\,$, black), no REM phase (\PnAD , pink) and \PAD $\,$ with a REM phase using a single episodic memory only (`w/o memory mix', purple).
    \textbf{(c, d)} Classification accuracy obtained on test datasets (c: CIFAR-10; d: SVHN) after training the linear classifier to convergence on the latent space $\boldsymbol{z}$ for each epoch of the $E$-$G$-network learning.
    Full model (\PAD): black line; without REM (\PnAD): pink line; with REM, but without memory mix: purple line.
    Solid lines represent mean and shaded areas indicate $\pm 1$ SEM over 4 different initial conditions.
  }\label{fig:results_linear_REM}}
\end{figure}

Semantic knowledge is fundamental for animals to learn quickly, adapt to new environments and communicate, and is hypothesized to be held by so-called semantic representations in cortex \citep{dicarlo_how_2012}. An example of such semantic representations are neurons from higher visual areas that contain linearly separable information about object category, invariant to other factors of variation, such as background, orientation or pose \citep{grill-spector_2001_the, hung_fast_2005, majaj_simple_2015}.

Here we demonstrate that \PAD, due to the specific combination of plasticity mechanisms during Wake, NREM and REM, develops such semantic representations in higher visual areas.
Similarly as in the previous section, we train our model on the CIFAR-10 and SVHN datasets.
To quantify the quality of inferred latent representations, we measure how easily downstream neurons can read out object identity from these.
For a simple linear read-out, its classification accuracy reflects the linear separability of different contents represented in a given dataset.
Technically, we train a linear classifier that distinguishes object categories based on their latent representations $\z$ after different numbers of wake-sleep cycles (`epochs', Fig.~\ref{fig:results_linear_REM}a) and report its accuracy on data not used during training of the model and classifier (``test data'').
While training the classifier, the connectivity of the network ($E$ and $G$) is fixed.

The latent representation ($\z$) emerging from the trained network
(Fig.~\ref{fig:results_linear_REM}b, full model) shows increasing linear separability reaching around 59\% test accuracy on CIFAR-10 (Fig.~\ref{fig:results_linear_REM}c, black line, for details see Supplements Table~\ref{tab:linear}) and 79\% on SVHN (Fig.~\ref{fig:results_linear_REM}d, black line), comparable to less biologically plausible machine-learning models \citep{berthelot_understanding_2018}.
These results show the ability of \PAD $\,$ to discover semantic concepts across wake-sleep cycles in an unsupervised fashion. 

Within our computational framework, we can easily consider sleep pathologies by directly interfering with the sleep phases.
To highlight the importance of REM in learning semantic representations, we consider a reduced model (\PnAD) in which the REM phase with adversarial dreaming is suppressed and only perturbed dreaming during NREM remains (Fig.~\ref{fig:results_linear_REM}b, pink cross).
Without REM sleep, linear separability increases much slower and even after a large number of epochs remains significantly below the \PAD $\,$ (see also Supplements Fig.~\ref{fig:supp_linear}c,d).
This suggests that adversarial dreaming during REM, here modeled by an adversarial game between feedforward and feedback pathways, is essential for the emergence of easily readable, semantic representations in the cortex.
From a computational point of view, this result is in line with previous work showing that learning to generate virtual inputs via adversarial learning (GANs variants) forms better representations than simply learning to reproduce external inputs \citep{radford_unsupervised_2015, donahue_adversarial_2016, berthelot_understanding_2018}. 

Finally, we consider a different pathology in which REM is not driven by randomly combined episodic memories and noise, but by single episodic memories without noise, as during NREM (Fig.~\ref{fig:results_linear_REM}b, purple cross).
Similarly to removing REM, linear separability increases much slower across epochs, leading to worse performance of the readout (Fig.~\ref{fig:results_linear_REM}c,d, purple lines).
For the SVHN dataset, the performance does not reach the level of the \PAD $\,$ even after many wake-sleep cycles (see also Supplements Fig.~\ref{fig:supp_linear}d).
This suggests that combining different, possibly non-related episodic memories, together with spontaneous cortical activity, as reported during REM dreaming \citep{fosse_dreaming_2003}, leads to significantly faster representation learning.

Our results suggest that generating virtual sensory inputs during REM dreaming, via a high-level combination of hippocampal memories and spontaneous cortical activity and subsequent adversarial learning,  allow animals to extract semantic concepts from their sensorium.
Our model provides hypotheses about the effects of REM deprivation, complementing pharmacological and optogenetic studies reporting impairments in the learning of complex rules and spatial object recognition \citep{boyce_causal_2016}.
For example, our model predicts that object identity would be less easily decodable from recordings of neuronal activity in the Inferior-Temporal (IT) cortex in animal models with chronically impaired REM sleep.

\subsection{Perturbed dreaming during NREM improves robustness of semantic representations.}

\begin{figure}[tbp!]
	{\centering
	\includegraphics[width=0.9\textwidth]{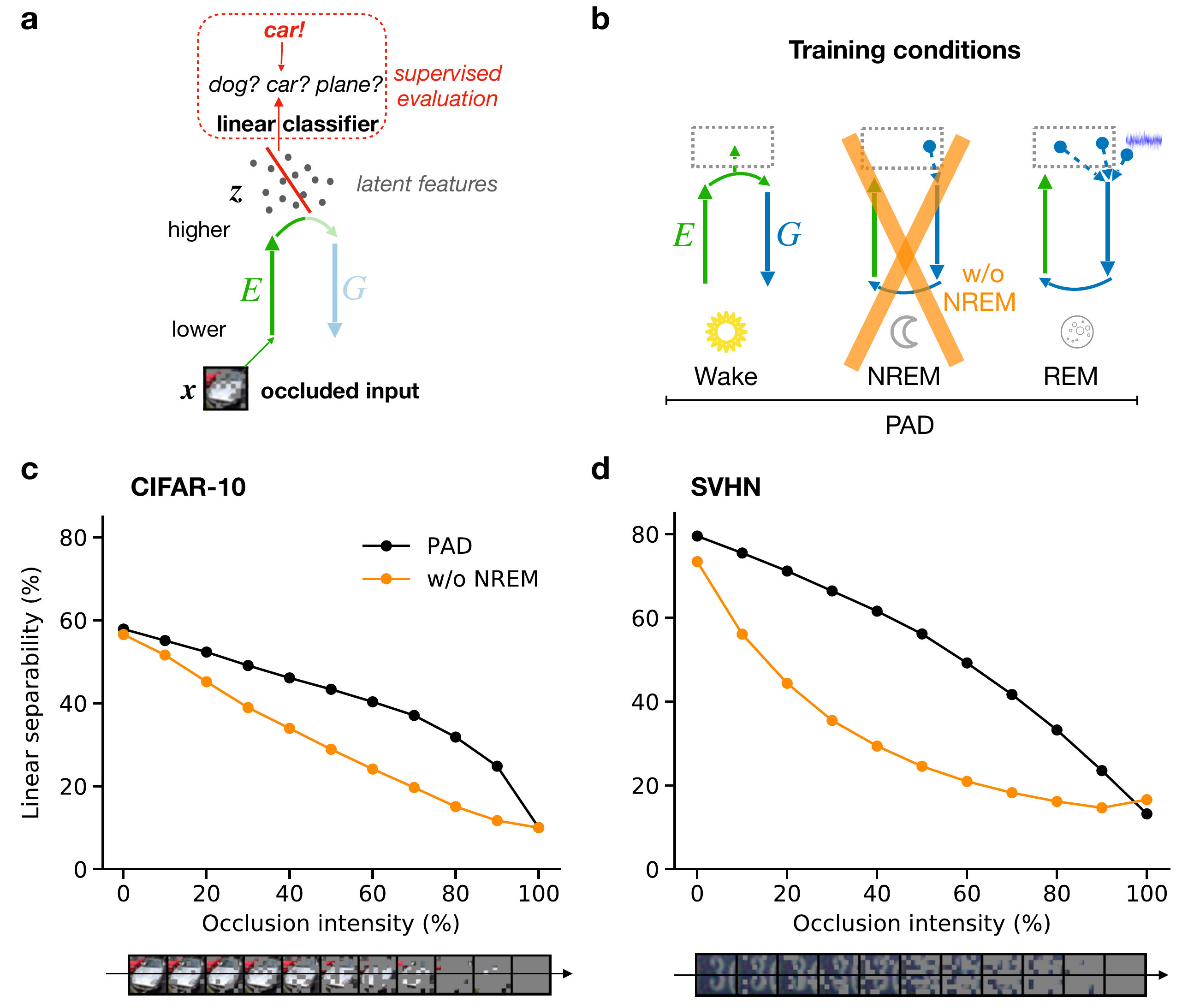}
	\caption{{\bf Perturbed dreaming during NREM improves robustness of latent representations.}
	\textbf{(a)} A trained linear classifier (cf. Fig.~\ref{fig:results_linear_REM}) infers class labels from latent representations.
	The classifier was trained on latent representations of original images, but evaluated on representations of images with varying levels of occlusion.
	\textbf{(b)} Training phases and pathological conditions: full model (\PAD, black), without NREM phase (\nPAD, orange).
	\textbf{(c, d)} Classification accuracy obtained on test dataset (C: CIFAR-10; D: SVHN) after $50$ epochs for different levels of occlusion (0 to 100\%).
	Full model (\PAD): black line; w/o NREM  (\nPAD): orange line. SEM over 4 different initial conditions overlap with data points.
	Note that due to an unbalanced distribution of samples the highest performance of a naive classifier is $18.9\%$ for the SVHN dataset.
	}\label{fig:results_linear_NREM}}
\end{figure}

Generalizing beyond previously experienced stimuli is essential for an animal's survival.
This generalization is required due to natural perturbations of sensory inputs, for example partial occlusions, noise, or varying viewing angles.
These alter the stimulation pattern, but in general should not change its latent representation subsequently used to make decisions.

Here, we model such sensory perturbations by silencing patches of neurons in early sensory areas during the stimulus presentation (Fig.~\ref{fig:results_linear_NREM}a).
As before, linear separability is measured via a linear classifier that has been trained on latent representations of un-occluded images and we use stimuli which were not used during training.
Adding occlusions hence directly tests the out-of-distribution generalization capabilities of the learned representations.
For the model trained with all phases (Fig.~\ref{fig:results_linear_NREM}b, full model), the linear separability of latent representations decreases as occlusion intensity increases, until reaching chance level for fully occluded images (Fig.~\ref{fig:results_linear_NREM}c,d; black line).

We next consider a sleep pathology in which we suppress perturbed dreaming during the NREM phase while keeping adversarial dreaming during REM (\nPAD, Fig.~\ref{fig:results_linear_NREM}B, orange cross).
In \nPAD, linear separability of partially occluded images is significantly decreased for identical occlusion levels (Fig.~\ref{fig:results_linear_NREM}c, d; compare black and orange lines).
In particular, performance degrades much faster with increasing occlusion levels. 
Note that despite the additional training objective, the full \PAD $\,$ develops equally good or even better latent representations of unoccluded images (0\% occlusion intensity) compared to this pathological condition without perturbed dreams. 

Crucially, the perturbed dreams in NREM are generated by replaying single episodic memories.
If the latent activity fed to the generator during NREM was of similar origin as during REM, i.e. obtained from a convex combination of multiple episodic memories coupled with cortical spontaneous activity, the quality of the latent representations significantly decreases (see also Supplements Fig.~\ref{fig:supp_nrem_mix}).
This suggests that only replaying single memories, as hypothesized to occur during NREM sleep \citep{oneill_play_2010}, rather than their noisy combination, is beneficial to robustify latent representations against input perturbations.

This robustification originates from the training objective defined in the NREM phase, forcing feedforward pathways to map perturbed inputs to the latent representation corresponding to their clean, non-occluded version. 
This procedure is reminiscent of a regularization technique from machine learning called `data augmentation' \citep{shorten2019survey}, which increases the amount of training data by adding stochastic perturbations to each input sample.
However, in contrast to data augmentation methods which directly operate on samples, here the system autonomously generates augmented data in offline states, preventing interference with online cognition and avoiding storage of the original samples.
Our `dream augmentation' suggests that NREM hippocampal replay not only maintains or strengthens cortical memories, as traditionally suggested \citep{klinzing_mechanisms_2019}, but also improves latent representations when only partial information is available.
For example, our model predicts that animals lacking such dream augmentation, potentially due to impaired NREM sleep, fail to react reliably to partially occluded stimuli even though their responses to clean stimuli are accurate.

\subsection{Latent organization in healthy and pathological models}

\begin{figure}[t]
	{\centering
	\includegraphics[width=1.0\textwidth]{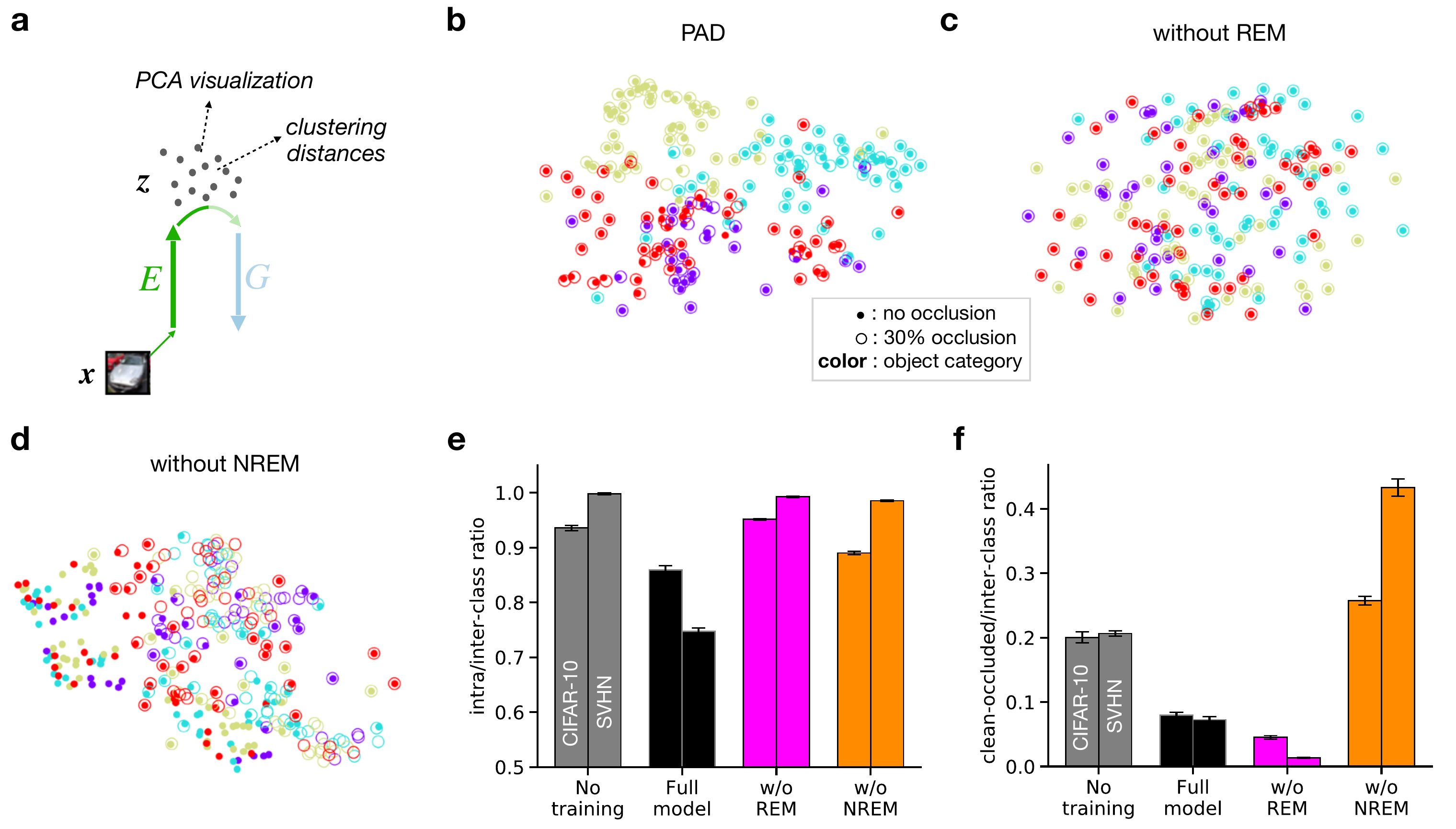}
	\caption{{\bf Effects of NREM and REM sleep on latent representations.}
	{\bf (a)} Inputs $\x$ are mapped to their corresponding latent representations $\z$ via the encoder $E$.
    Principal Component Analysis \citep{jolliffe_principal_2016} is performed on the latent space to visualize its structure (b-d).
    Clustering distances (e,f) are computed directly on latent features $\z$.
	{\bf (b, c, d)} 
	PCA visualization of latent representations projected on the first two principal components.
	Full circles represent clean images, open circles represent images with $30\%$ occlusion. Each color represents an object category from the SVHN dataset (purple:`0', cyan:`1', yellow:`2', red:`3').
	{\bf (e)}  Ratio between average intra-class and average inter-class distances in latent space for randomly initialized networks (no training, grey), full model (black), model trained without REM sleep (w/o REM, pink) and model trained without NREM sleep (w/o NREM, orange) for un-occluded inputs.
	{\bf (f)} Ratio between average clean-occluded (30\% occlusion) and average inter-class distances in latent space for full model (black), w/o REM (pink) and w/o NREM (orange). Error bars represent SEM over 4 different initial conditions.
	}\label{fig:results_PCA}}
\end{figure}

The results so far demonstrate that perturbed and adversarial dreaming (\PAD), during REM and NREM sleep states, contribute to cortical representation learning by increasing the linear separability of latent representations into object classes.
We next investigate how the learned latent space is organized, i.e., whether representations of sensory inputs with similar semantic content are grouped together even if their low-level structure may be quite different, for example due to different viewing angles, variations among an object category, or (partial) occlusions.

We illustrate the latent organization by projecting the latent variable $\z$ using 
Principal Component Analysis \citep[PCA, Fig.~\ref{fig:results_PCA}a,][]{jolliffe_principal_2016}. This method is well-suited for visualizing high-dimensional data in a low-dimensional space while preserving as much of the data's variation as possible.

For \PAD, the obtained PCA projection shows relatively distinct clusters of latent representations according to the semantic category ("class identity") of their corresponding images (Fig.~\ref{fig:results_PCA}b).
The model thus tends to organize latent representations such that high-level, semantic clusters are discernable.
Furthermore, partially occluded objects (Fig.~\ref{fig:results_PCA}b, empty circles) are represented closeby their corresponding un-occluded version (Fig.~\ref{fig:results_PCA}b, full circles). 

As shown in the previous sections, removing either REM or NREM has a negative impact on the linear separability of sensory inputs.
However, the reasons for these effects are different between REM and NREM.
If REM sleep is removed from training (\PnAD), representations of unoccluded images are less organized according their semantic category, but still match their corresponding occluded versions (Fig.~\ref{fig:results_PCA}c).
REM is thus necessary to organize latent representations into semantic clusters, providing an easily readable representation for downstream neurons.
In contrast, removing NREM (\nPAD) causes representations of occluded inputs to be remote from their un-occluded representations (Fig.~\ref{fig:results_PCA}d).

We quantify these observations by computing the average distances between latent representations from the same object category (intra-class distance) and between representations of different object category (inter-class distance).
Since the absolute distances are difficult to interpret, we focus on their ratio (Fig.~\ref{fig:results_PCA}e).
On both datasets, this ratio increases if the REM phase is removed from training (Fig.~\ref{fig:results_PCA}e, compare black and pink bars), reaching levels comparable to the one with the untrained network.
Moreover, removing NREM from training also increases this ratio.
These observations suggest that both perturbed and adversarial dreaming jointly reorganize the latent space such that stimuli with similar semantic structure are mapped to similar latent representations.
In addition, we compute the distance between the latent representations inferred from clean images and their corresponding occluded versions, also divided by the inter-class distance (Fig.~\ref{fig:results_PCA}f).
By removing NREM from training, this ratio increases significantly, highlighting the importance of NREM in making latent representations invariant to input perturbations.

\subsection{Cortical implementation of PAD}

We have shown that perturbed and adversarial dreaming (\PAD) can learn semantic cortical representations useful for downstream tasks.
Here we hypothesize how the associated mechanisms may be implemented in cortex. 

First, PAD implies the existence of discriminator neurons that would learn to be differentially active during wakefulness and REM sleep.
It also postulates a conductor that orchestrates learning by providing a teaching (`nudging') signal to the discriminator neurons during Wake and REM.
Experimental evidence suggests that discriminator neurons, differentiating between internally generated end externally driven sensory activity, may reside in the anterior cingulate cortex (ACC) or the medial prefrontal cortex (mPFC), but functionally similar neurons may be located across cortex to deliver local learning signals \citep{Subramaniam2012, simons_brain_2017,gershman_generative_2019,benjamin_learning_2021}.

\begin{figure}[t!]
  {\centering
    \includegraphics[width=0.5\textwidth]{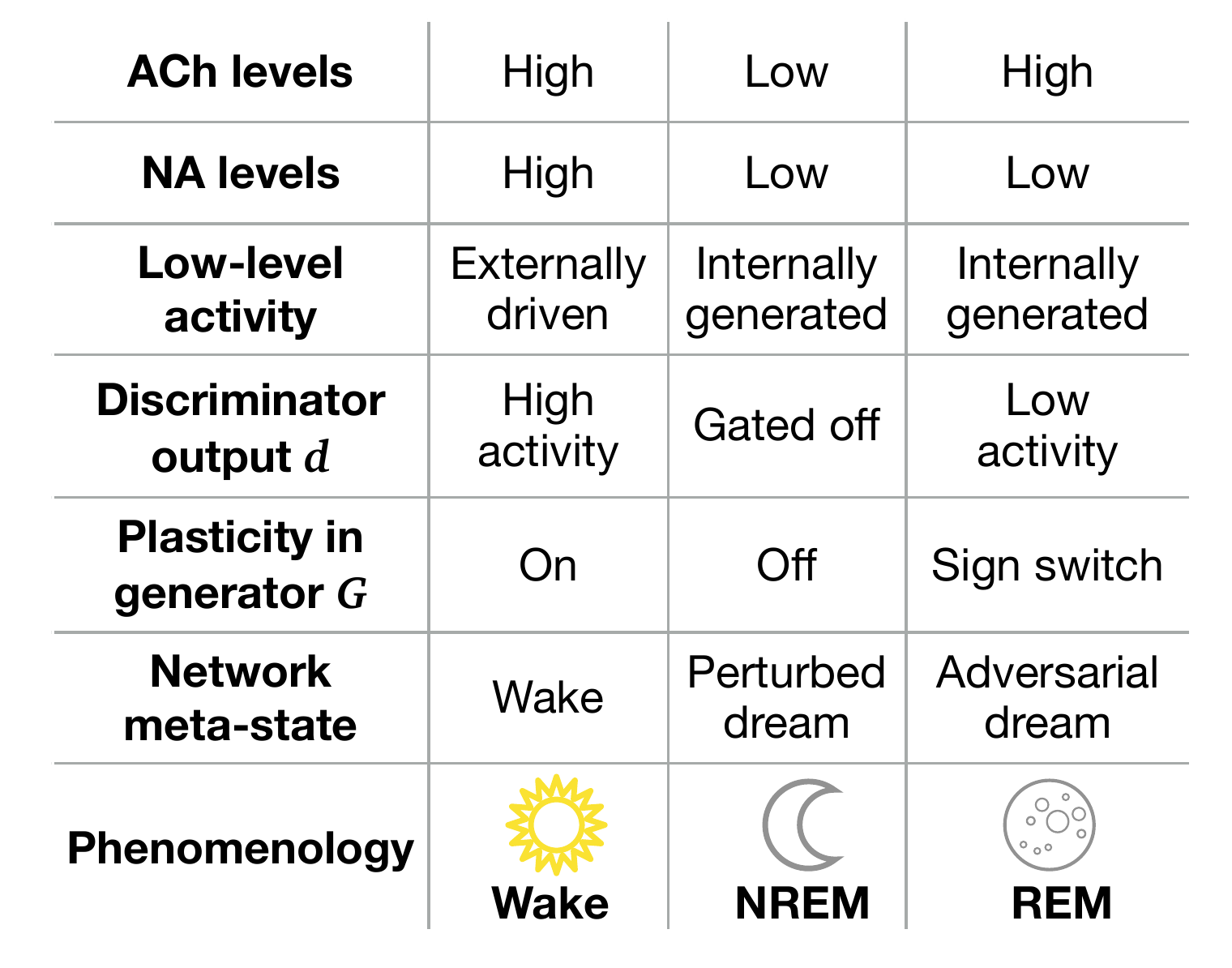}
    \caption{{\bf Model features and physiological counterparts during Wake, NREM and REM phases.}
      ACh: acetylcholine; NA: noradrenaline.
      "Sign switch" indicates that identical local errors lead to opposing weight changes between Wake and REM sleep.
      }\label{fig:results_cortical}}
\end{figure}

Second, learning in PAD is orchestrated across three different phases: (i) learning stimulus reconstruction during Wake, (ii) learning latent variable reconstruction during NREM sleep (’perturbed dreaming’), and (iii) learning to generate realistic sensory activity during REM sleep (’adversarial dreaming’).
Our model suggests that objective functions and synaptic plasticity are affected by these phases (Fig.~\ref{fig:results_cortical}).
Wakefulness is associated with increased activity of modulatory brainstem neurons releasing neuromodulators such as acetylcholine (ACh) and noradrenaline (NA), hypothesized to prioritize the amplification of information from external stimuli \citep{Adamantidis2019,aru_apical_2020}.
In contrast, neuromodulator concentrations during NREM are reduced compared to Wake, while REM is characterized by high ACh and low NA levels \citep{hobson_rem_2009}.
We postulate that the state-specific modulation provides a high activity target for the discriminator during Wake which is decreased during REM and entirely gated off during NREM.
Furthermore, we suggest that adversarial learning is implemented by a sign-switched plasticity in the generative network during REM sleep, with respect to Wake.
During wakefulness, plasticity in these apical synapses may be enhanced by noradrenaline (NA) as opposed to NREM \citep{Adamantidis2019,aru_apical_2020}.
The presence of acetylcholine (ACh) alone during REM \citep{hobson_dreaming_2000} may switch the sign of plasticity in apical synapses of (hippocampal) pyramidal neurons \citep{McKay2007}.
Furthermore, it is known that somato-dendritic synchrony is reduced in REM versus NREM sleep \citep{Seibt2017}; this suggests a reduced somato-dendritic backpropagation of action potentials, which, in turn, is known to switch the sign of apical plasticity \citep{Sjostrom2006a}.

Third, learning in our model requires the computation of reconstruction errors, i.e., mismatches between top-down and bottom-up activity.
So far, two non-exclusive candidates for computing mismatch signals have been proposed.
One suggests a dendritic error representation in layer 5 pyramidal neurons that compare bottom-up with top-down inputs from our encoding ($E$) and generative ($G$) pathways \citep{guerguiev_towards_2017, sacramento_dendritic_2017}.
The other suggests an explicit mismatch representation by subclasses of layer 2/3 pyramidal neurons \citep{Keller2018}.

Fourth, our computational framework assumes effectively separate feedforward and feedback streams.
A functional separation of these streams does not necessarily imply a structural separation at the network level. Indeed, such cross-projections are observed in experimental data \citep{gilbert_top-down_2013} and also used in, e.g., the predictive processing framework \citep{rao_predictive_1999}.
In our model, an effective separation of the information flows is required to prevent "information shortcuts" across early sensory cortices which would prevent learning of good representations in higher sensory areas.
This suggests that for significant periods of time, intra-areal lateral interactions between cortical feedforward and feedback pathways are effectively gated off in most of the areas. 

Fifth, similar to previous work \citep{kali_off-line_2004}, the hippocampus is not explicitly modeled but rather mimicked by a buffer allowing simple store and retrieve operations.
An extension of our model could replace this simple mechanism with attractor networks which have been previously employed to model hippocampal function \citep{tang_memory_2010}.
The combination of episodic memories underlying REM dreams in our model could either occur in hippocampus or in cortex.
In either case, we would predict a nearly simultaneous activation of different episodic memories in hippocampus that results in the generation of creative virtual early cortical activity.

Finally, beyond the mechanisms discussed above, our model assumes that cortical circuits can efficiently perform credit assignment, similar to the classical error backpropagation algorithm.
Most biologically plausible implementations for error-backpropagation involve feedback connections to deliver error signals \citep{whittington_theories_2019, richards_2019, lillicrap_backpropagation_2020}, for example to the apical dendrites of pyramidal neurons \citep{sacramento_dendritic_2017, guerguiev_towards_2017, haider2021latent}. 
An implementation of our model in such a framework would hence require additional feedforward and feedback connections for each neuron.
For example, neurons in the feedforward pathway would not only project to higher cortical areas to transmit signals, but additionally project back to earlier areas to allow these to compute the local errors required for effective learning.
Overall, our proposed model could be mechanistically implemented in cortical networks through different classes of pyramidal neurons with a biological version of supervised learning based on a dendritic prediction of somatic activity  \citep{Urbanczik2014}, and a corresponding global modulation of synaptic plasticity by state-specific neuromodulators.


\section{Discussion}

Semantic representations in cortical networks emerge in early life despite most observations lacking an explicit class label, and sleep has been hypothesized to facilitate this process \citep{klinzing_mechanisms_2019}.
However, the role of dreams in cortical representation learning remains unclear.
Here we proposed that creating virtual sensory experiences by randomly combining episodic memories during REM sleep lies at the heart of cortical representation learning.
Based on a functional cortical architecture, we introduced the perturbed and adversarial dreaming model (PAD) and demonstrated that REM sleep can implement an adversarial learning process which, constrained by the network architecture and the choice of latent prior distributions, builds semantically organized latent representations.
Additionally, perturbed dreaming based on the episodic memory replay during NREM stabilizes the cortical representations against sensory perturbations.
Our computational framework allowed us to investigate the effects of specific sleep-related pathologies on cortical representations.
Together, our results demonstrate complementary effects of perturbed dreaming from individual episodes during NREM and adversarial dreaming from mixed episodes during REM.
PAD suggests that the generalization abilities exhibited by humans and other animals arise from distinct processes during the two sleep phases: REM dreams organize representations semantically and NREM dreams stabilize these representations against perturbations.
Finally, the model suggests how adversarial learning inspired by GANs can potentially be implemented by cortical circuits and associated plasticity mechanisms. 
 
\subsection*{Relation to cognitive theories of sleep}

PAD focuses on the functional role of sleep, and in particular dreams.
Many dynamical features of brain states during NREM and REM sleep, such as cortical oscillations \citep{leger_slow-wave_2018} are hence ignored here but will potentially become relevant when constructing detailed circuit models of the suggested architectures, for example for switching between memories \citep{korcsakgorzo2021cortical}.
Our proposed model of sleep is complementary to theories suggesting that sleep is important for physiological and cognitive maintenance \citep{mcclelland_why_1995, kali_off-line_2004, renno-costa_computational_2019, van_de_ven_brain-inspired_2020}.
In particular, \citet{norman_methods_2005} proposed a model where autonomous reactivation of memories (from cortex and hippocampus) coupled with oscillating inhibition during REM sleep helps detect weak parts of memories and selectively strengthen them, to overcome catastrophic forgetting.
While our REM phase serves different purposes, an interesting commonality is the view of REM as a period where the cortex "thinks about what it already knows" from past and recent memories and reorganizes its representations by replaying them together, as opposed to NREM  where only recent memories are replayed and consolidated.
Recent work has also suggested that the brain learns using adversarial principles, either as a reality monitoring mechanism potentially explaining delusions in some mental disorders \citep{gershman_generative_2019}, in the context of dreams to overcome overfitting and promote generalization \citep{hoel_overfitted_2021}, and for learning inference in recurrent biological networks \citep{benjamin_learning_2021}.

Cognitive theories propose that sleep promotes the abstraction of semantic concepts from episodic memories through a hippocampo-cortical replay of waking experiences, referred to as "memory semantization" \citep{nadel_memory_1997, lewis_overlapping_2011}.
The learning of organized representations is an important basis for semantization.
An extension of our model would consider the influence of different sensory modalities on representation learning \citep{guo_multimodal_2019}, which is known to significantly influence cortical schemas \citep{lewis_how_2018} and can encourage the formation of computationally powerful representations \citep{radford2021learning}. 

Finally, sleep has previously been considered as a state where `noisy' connections acquired during wakefulness are selectively forgotten \citep{Crick1983, Poe2017}, or similarly, as a homeostatic process to desaturate learning and renormalize synaptic strength \citep[synaptic homeostasis hypothesis; ][]{tononi_sleep_2014, Tononi2020}. 
In contrast, our model offers an additional interpretation of plasticity during sleep, where synapses are globally readapted to satisfy different but complementary learning objectives than Wake, either by improving feedforward recognition of perturbed inputs (NREM) or by adversarially tuning top-down generation (REM).

\subsection*{Relation to representation learning models}

Recent advances in machine learning, such as self-supervised learning approaches, have provided powerful techniques to extract semantic information from complex datasets \citep{liu_self-supervised_2021}.
Here, we mainly took inspiration from self-supervised generative models combining autoencoder and adversarial learning approaches \citep{radford_unsupervised_2015, donahue_adversarial_2016, dumoulin_adversarially_2017, berthelot_understanding_2018, liu_self-supervised_2021}.
It is theoretically not yet fully understood how linearly separable representations are learned from objectives which do not explicitly encourage them, i.e., reconstruction and adversarial losses.
We hypothesize that the presence of architectural constraints and latent priors, in combination with our objectives, enable their emergence \citep[see also][]{alemi2018fixing,tschannen2020mutual}.
Note that similar generative machine learning models often report a higher linear separability of network representations, but use all convolutional layers as a basis for the readout \citep{radford_unsupervised_2015, dumoulin_adversarially_2017}, while we only used low-dimensional features $\z$.
Approaches similar to ours, i.e., those which perform classification only on the latent features, report comparable performance to ours \citep{berthelot_understanding_2018, hjelm_learning_2019, beckham_adversarial_nodate}.

Furthermore, in contrast to previous GAN variants, our model removes many optimization tricks such as batch-normalization layers \citep{loffe_batch_2018}, spectral normalization layers \citep{miyato_spectral_2018} or optimizing the min-max GAN objective in three steps with different objectives, which are challenging to implement in biological substrates.
Despite their absence, our model maintains a high quality of latent representations.
As our model is relatively simple, it is amenable to implementations within frameworks approximating backpropagation in the brain \citep{whittington_theories_2019, richards_2019, lillicrap_backpropagation_2020}.  
However, some components remain challenging for implementations in biological substrates, for example convolutional layers \citep[but see][]{pogodin_towards_2021} and batched training \citep[but see][]{marblestone2016toward}.

\subsection*{Dream augmentations, mixing strategies and fine-tuning}

To make representations robust, a computational strategy consists of learning to map different sensory inputs containing the same object to the same latent representation, a procedure reminiscent of data augmentation \citep{shorten2019survey}.
As mentioned above, unlike standard data augmentation methods, our NREM phase does not require the storage of raw sensory inputs to create altered inputs necessary for such data augmentation and instead relies on (hippocampal) replay being able to regenerate similar inputs from high-level representations stored during wakefulness.
Our results obtained through perturbed dreaming during NREM provide initial evidence that this dream augmentation may robustify cortical representations.

Furthermore, as discussed above, introducing more specific modifications of the replayed activity, for example mimicking translations or rotations of objects, coupled with a negative phase where latent representations from different images are pushed apart, may further contribute to the formation of invariant representations.
Along this line, recent self-supervised contrastive learning methods \citep{gidaris_unsupervised_2018, chen_simple_2020, zbontar2021barlow} have been shown to enhance the semantic structure of latent representations by using a similarity objective where representations of stimuli under different views are pulled together in a first phase, while, crucially, embedding distances between unrelated images are increased in a second phase.

In our REM phase, different mixing strategies in the latent layer could be considered. 
For instance, latent activities could be mixed up by retaining some vector components of a representation and using the rest from a second one \citep{beckham_adversarial_nodate}.
Moreover, more than two memory representations could have been used. Alternatively, our model could be trained with spontaneous cortical activity only.
In our experimental setting we do not observe significant differences between using a combination of episodic memories with spontaneous activity or only using spontaneous activity (Supplements Fig.~\ref{fig:supp_linear_gaussian}). However, we hypothesize that for models which learn continuously, a preferential replay of combinations of recent episodic memories encourages the formation of cortical representations that are useful in the present.

Here, we used a simple linear classifier to measure the quality of latent representations, which is an obvious simplification with regard to cortical processing.
Note however that also for more complex `readouts', organized latent representations enable more efficient and faster learning \citep{silver2017predictron,ha2018world,schrittwieser2020mastering}.
In its current form, PAD assumes that training the linear readout does not lead to weight changes in the encoder network.
However, in cortical networks, cognitive or motor tasks leveraging latent representations likely shape the encoder network, which could in our model be reflected in `fine-tuning' the encoder for specific tasks \citep[compare][]{liu_self-supervised_2021}.

Finally, our model does not show significant differences in performance when the order of sleep phases is switched (Supplements Fig.~\ref{fig:supp_linear_order}).
However, NREM and REM are observed to occur in a specific order throughout the night \citep{diekelmann_memory_2010} and this order has been hypothesized to be important for memory consolidation \citep["sequential hypothesis", ][]{giuditta1998}.
The independence of phases in our model may be due to the relatively small synaptic changes occurring in each phase.
We expect the order of sleep phases to influence model performance if these changes become larger, either due to longer phases or increased learning rates. The latter may become particularly relevant in continual learning settings where it becomes important to control the emphasis put on recent observations.

\subsection*{Signatures of generative learning}

\PAD$\,$ makes several experimentally testable predictions at the neuronal and systems level.
We first address generally whether the brain learns via generative models during sleep before discussing specific signatures of adversarial learning.

First, our NREM phase assumes that hippocampal replay generates perturbed wake-like early sensory activity \citep[see also][]{ji_coordinated_2007} which is subsequently processed by feedforward pathways.
Moreover, our model predicts that over the course of learning, sensory-evoked neuronal activity and internally-generated activity during sleep become more similar.
In particular, we predict that (spatial) activity in both NREM and REM become more similar to Wake, however, patterns observed during REM remain distinctly different due to the creative combination of episodic memories.
Future experimental studies could confirm these hypotheses by recording early sensory activity during wakefulness, NREM and REM sleep at different developmental stages and evaluating commonalities and differences between activity patterns.
Previous work has already demonstrated increasing similarity between stimulus-evoked and spontaneous (generated) activity patterns during wakefulness in ferret visual cortex  (\citealp{berkes2011spontaneous}; but see \citealp{avitan2021spontaneous}). 

On a behavioral level, the improvement of internally-generated activity patterns correlates with the development of dreams in children, that are initially unstructured, simple and plain, and gradually become full-fledged, meaningful, narrative, implicating known characters and reflecting life episodes \citep{nir_dreaming_2010}.
In spite of their increase in realism, REM dreams in adulthood are still reported as bizarre \citep{williams_bizarreness_1992}.
Bizarre dreams, such as a ``flying dogs'', are typically defined as discontinuities or incongruities of the sensory experience \citep{mamelak_dream_1989} rather than completely structureless experiences.
This definition hence focuses on high-level logical structure, not on the low-level sensory content.
In contrast, the low FID score, i.e., high realism, of REM dreams in our experiments reflects that the low-level structure on which this evaluation metric mainly focuses \citep[e.g.,][]{brendel2019approximating} is similar to actual sensory input.
Capturing the "logical realism" of our generated neuronal activities most likely requires a more sophisticated evaluation metric and an extension of the model capable of generating temporal sequences of sensory stimulation.
We note, however, that even such surreal dreams as ``flying dogs'' can be interpreted as altered combinations of episodic memories and thus, in principle, can arise from our model.

Second, our model suggests that the development of semantic representations is mainly driven by REM sleep.
This allows us to make predictions which connect the network with the systems level, in the specific case of acquiring skills from complex and unfamiliar sensory input.
For humans, this could be learning a foreign language with unfamiliar phonetics.
Initially, cortical representations cannot reflect relevant nuances in these sounds.
Phonetic representations develop gradually over experience and are reflected in changes of the sensory evoked latent activity, specifically in the reallocation of neuronal resources to represent the relevant latent dimensions.
We hypothesize that in case of impaired REM sleep, this change of latent representations is significantly reduced, which goes hand in hand with decreased learning speed.
Future experimental studies could investigate these effects for instance by trying to decode sound identity from high-level cortical areas in patients where REM sleep is impaired over long periods through pharmacological agents such as anti-depressants \citep{boyce_rem_2017}.
An equivalent task in the non-human animal domain would be song acquisition in songbirds \citep{fiete_birdsong_2007}.
On a neuronal level, one could selectively silence feedback pathways during REM sleep in animal models over many nights, for example via optogenetic tools.
Our model predicts that this silencing would significantly impact the animal's learning speed, as reported from animals with reduced theta rhythm during REM sleep \citep{boyce_rem_2017}.

\subsection*{Signatures of adversarial learning}

The experimental predictions discussed above mainly address whether the brain learns via generative models during sleep.
Here we make experimental predictions which would support our hypotheses and contrast it to alternative theories of learning during offline states.

\paragraph{Existence of an external/internal discriminator} 
The discriminator provides our model with the ability to distinguish externally driven from internally generated low-level cortical activity.
Due to this unique property, the discriminator may be leveraged to distinguish actual from imagined sensations.
According to our model, reduced REM sleep would lead to an impaired discriminator, and could thus result in an inability of subjects to realize that self-generated imagery is not part of the external sensorium.
This may result in the formation of delusions, as previously suggested \citep{gershman_generative_2019}.
For instance, hallucinations in schizophrenic patients, often mistaken for veridical perceptions \citep{waters_what_2016}, could be partially caused by abnormal REM sleep patterns, related to observed reduced REM latency and density \citep{cohrs_sleep_2008}.
Based on these observations, we predict in the context of our model a negative correlation between REM sleep quality and delusional perceptions of hallucinations.
Systematic differences in REM sleep quality may hence explain why some patients are able to recognize that their hallucinations are self-generated while some others mistake them to be real.
Moreoever, although locating discriminator neurons may prove non-trivial (but see "Cortical implementation" for specific suggestions), we predict that once the relevant cells have been identified, perturbing them may lead to detrimental effects on differentiating between external sensory inputs and internally generated percepts.

The state-specific activity of the discriminator population makes predictions about plasticity on synapses in the feedforward stream during wakefulness and sleep.
In our model, the discriminator is trained to distinguish externally from internally generated patterns by opposed targets imposed during Wake and REM.
After many wake-sleep cycles, the KL loss as well as the reconstruction loss (see Methods) in our model become small compared to the adversarial loss (Fig.~\ref{fig:supp_losses_cifar10}, \ref{fig:supp_losses_svhn}), which remains non-zero due to a balance between discriminator and generator.
The same low-level activity pattern would hence cause opposite weight changes during wakefulness and sleep on feedforward synapses.
This could be tested experimentally by actively instantiating similar spatial activity patterns in low-level sensory cortex during wakefulness and REM and compare the statistics of observed changes in (feedforward) downstream synapses.

\paragraph{Adversarial training of a generator during sleep} 
To drive adversarial learning and maintain a balance between the generator and discriminator, the generative network must be trained in parallel to the discriminative (encoder) network during REM.
In contrast, in alternative representation learning models which involve offline states such as the Wake-Sleep algorithm \citep{hinton_wake-sleep_1995}, generative pathways are not trained to produce realistic dreams during the sleep phase. Rather, they are trained by reconstruction on real input data during the wake phase.
This allows an experimental distinction between our model and Wake-Sleep-like models: while our model predicts plasticity in both bottom-up and top-down pathways both during wake and during REM sleep, Wake-Sleep models alternate between training feedback and feedforward connections during online and offline states, respectively.

Previous work has developed methods to infer plasticity rules from neuronal activity \citep{Lim2015,Senn2015} or weight changes \citep{nayebi2020identifying}.
In the spirit of existing {\em in vivo} experiments, we suggest to optogenetically monitor and potentially modulate apical dendritic activities in cortical pyramidal neurons of mice during wakefulness and REM sleep \citep{Li2017,Voigts2020,Schoenfeld2022}.
From the statistics of the recorded dendritic and neuronal activity, the plasticity rules could be inferred and compared to the state-dependent rules suggested by our model, in particular to the predicted sign-switch of plasticity between wakefulness and REM sleep.

\paragraph{Adversarial learning and creativity}

Adversarial learning, for example in GANs, enables a form of creativity, reflected in their ability to generate realistic new data or to create semantically meaningful interpolations \citep{radford_unsupervised_2015, berthelot_understanding_2018, karras_style-based_2018}.
This creativity might be partly caused by the freedom in generating sensory activity that is not restricted by requiring good reconstructions, but is only guided by the internal/external judgment \citep{goodfellow_nips_2016}.
This is less constraining on the generator than direct reconstruction losses used in alternative models such as variational auto-encoders \citep{kingma2013autoencoding} or the Wake-Sleep algorithm \citep{hinton_wake-sleep_1995}.
We thus predict that REM sleep, here implementing adversarial learning, should boost creativity, as previously reported \citep{cai_rem_2009, llewellyn_crossing_2016, lewis_how_2018}.
Furthermore, we predict that REM sleep influences a subject's ability to visualize creative mental images, for instance associating non-obvious visual patterns from distinct memories.
For example, we predict that participants chronically lacking REM sleep would perform worse than control participants at a creative synthesis task \citep{palmiero_domain-specificity_2015}, consisting of combining different visual components into a new, potentially useful object.

\paragraph{Adversarial learning and lucid dreaming}

Finally, adversarial dreaming offers a theoretical framework to investigate neuronal correlates of normal versus lucid dreaming \citep{Dresler2012,Baird2019}. 
While in normal dreaming the internally generated activity is perceived as externally caused, in lucid dreaming it is perceived as what it is, i.e., internally generated.
We hypothesize that the "neuronal conductor" that orchestrates adversarial dreaming is also involved in lucid dreaming, by providing to the dreamer conscious access to the target "internal" that the conductor imposes during REM sleep. Our cortical implementation suggests that the neuronal conductor could
gate the discriminator teaching via apical activity of cortical pyramidal neurons.
The same apical dendrites were also speculated to be involved in conscious perception \citep{Takahashi2020}, dreaming \citep{aru_apical_2020}, and in representing the state and content of consciousness \citep{Aru2019}.

\medskip
\medskip 

Our model demonstrates that adversarial learning during wakefulness and sleep can provide significant benefits to extract semantic concepts from sensory experience.
By bringing insights from modern artificial intelligence to cognitive theories of sleep function, we suggest that cortical representation learning during dreaming is a creative process, orchestrated by brain-state-regulated adversarial games between separated feedforward and feedback streams.  
Adversarial dreaming may further be helpful to understand learning beyond the standard student-teacher paradigm.
By `seeing' the world from new perspectives every night, dreaming represents an active learning phenomenon, constantly improving our understanding, our creativity and our awareness.


\section{Methods}

\subsection{Network architecture}

\begin{figure}[!h]
  {\centering
    \includegraphics[width=0.6\textwidth]{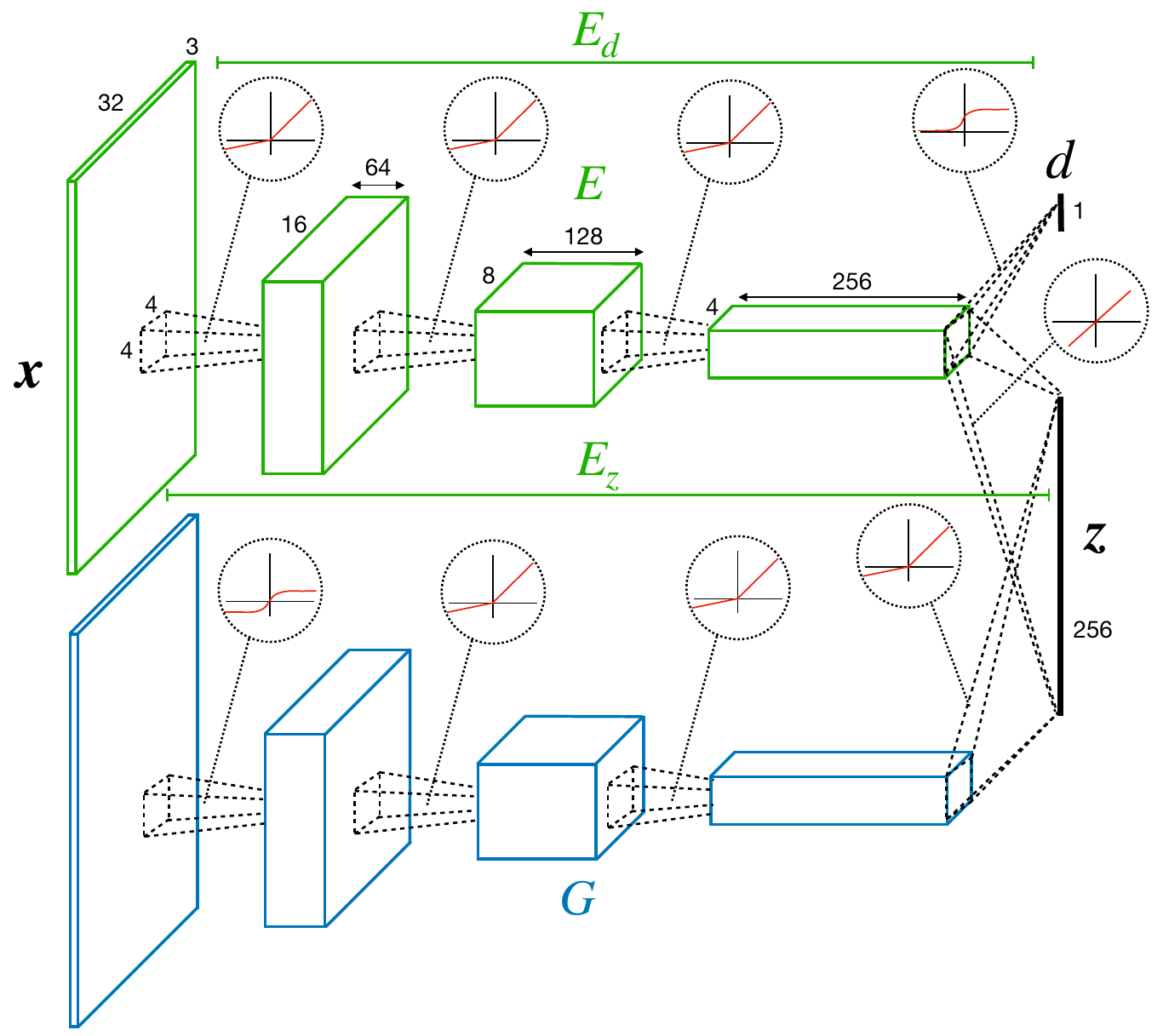}
    \caption{Convolutional neural network (CNN) architecture of encoder/discriminator and generator used in \PAD.
    }\label{fig:methods_CNN}
    \medskip}
  \small
\end{figure}

The network consists of two separate pathways, mapping from the pixel to the latent space (`encoder'/'discriminator') and from the latent to pixel space (`generator').
Encoder/Discriminator and Generator architectures follow a similar structure as the DCGANs model \citep{radford_unsupervised_2015}.
The encoder $E_z$ has four convolutional layers \citep{lecun_deep_2015} containing $64, 128, 256$ and $256$ channels respectively (Fig.~\ref{fig:methods_CNN}).
Each layer uses a $4\times 4$ kernel, a padding of $1$ ($0$ for last layer), and a stride of $2$, i.e., feature size is halved in each layer.
All convolutional layers except the last one are followed by a LeakyReLU non-linearity \citep{leakyrelu}.
We denote the activity in the last convolutional layer as $\z$.
An additional convolutional layer followed by a sigmoid non-linearity is added on top of the second-to-last layer of the encoder and maps to a single scalar value $d$, the internal/external discrimination (with putative teaching signal $0$ or $1$).
We denote the mapping from $\x$ to $d$ by $E_d$. $E_z$ and $E_d$ thus share the first three convolutional layers. We jointly denote them by $E$, where $E(\x) = (E_z(\x), E_d(\x)) = (\z, d)$ (Fig.~\ref{fig:methods_CNN}).

 Mirroring the structure of $E_z$, the generator $G$ has four deconvolutional layers containing $256, 128, 64$, and $3$ channels.
They all use a $4\times 4$ kernel, a padding of $1$ ($0$ for first deconvolutional layer) and a stride of $2$, i.e, the feature-size is doubled in each layer.
The first three deconvolutional layers are followed by a LeakyReLU non-linearity, and the last one by a tanh non-linearity.

As a detailed hippocampus model is outside the scope of this study, we mimic hippocampal storage and retrieval by storing and reading latent representations to and from memory.

\subsection{Datasets}
We use the CIFAR-10 \citep{cifar10} and SVHN \citep{svhn} datasets to evaluate our model.
They consist of $32\times 32$ pixel images with three color channels.
We consider their usual split into a training set and a smaller test set.

\subsection{Training procedure}
We train our model by performing stochastic gradient-descent with mini-batches on condition-specific objective functions, in the following also referred to as loss functions, using the ADAM-optimizer \citep[$\beta_1 = 0.5$, $\beta_2 = 0.999$;][]{kingma_adam_2017} with learning rate of 0.0002 and mini-batch size of 64.
We rely on our model being fully differentiable.
The following section describes the loss functions for the respective conditions.

\begin{algorithm}[!h]
\SetAlgoLined
 $\theta_E$, $\theta_G$ \tcp*{initialize network parameters}
 \For{number of training iterations}{
 \texttt{\\}
 Wake \\
  $\X \leftarrow \{\x^{(1)} , ..., \x^{(b)} \}$ \tcp*{random mini-batch from dataset}
  $\Z, \D \leftarrow E(\X) $ \tcp*{infer latent and discriminative outputs}
  $\X' \leftarrow G(\Z) $ \tcp*{reconstruct input via generator}
  $\mathcal{L}_{\text{img}} \leftarrow \frac{1}{b}\sum_{i=1}^b \| \x^{(i)} - {\x'}^{(i)} \|^2$ \tcp*{compute reconstruction loss}
  $\mathcal{L}_{\text{KL}} \leftarrow\text{D}_{\text{KL}} ( q(\Z) || p(\Z) )$ \tcp*{compute KL-loss}
   $\mathcal{L}_{\text{real}} \leftarrow  - \frac{1}{b} \sum_{i=1}^b \log(\disc^{(i)}) $ \tcp*{compute discriminator loss on real samples}
   $\theta_E \leftarrow \theta_E - \nabla_{\theta_E}(\mathcal{L}_{\text{img}} + \mathcal{L}_{\text{KL}} + \mathcal{L}_{\text{real}})  $ \tcp*{update encoder/discriminator parameters}
   $\theta_G \leftarrow \theta_G - \nabla_{\theta_G}\mathcal{L}_{\text{img}}  $ \tcp*{update generator parameters}

   \texttt{\\}
   NREM sleep \\
  $\Z \leftarrow \{\z^{(1)} , ..., \z^{(b)} \} $ \tcp*{mini-batch of latent vectors from Wake}
   $\X' \leftarrow G(\Z) $ \tcp*{reconstruct input via generator}
   $\Z' \leftarrow E_z(\X' \odot \boldsymbol{\Omega}) $ \tcp*{infer perturbed input}
   $\mathcal{L_{\text{NREM}}} \leftarrow \frac{1}{b} \sum_{i=1}^b \|\z^{(i)} - \z'^{(i)} \|^2$ \tcp*{compute reconstruction loss}
   $\theta_E \leftarrow \theta_E - \nabla_{\theta_E}\mathcal{L}_{\text{NREM}}$ \\

   \texttt{\\}
   REM sleep \\
  
  \eIf{first iteration}{
   $\Z_{\text{mix}} \leftarrow \Z$ \\
   }{
   $\Z_{\text{mix}} \leftarrow  \lambda' (\lambda \Z + (1 - \lambda) \Z_{\text{old}}) + (1-\lambda') \boldsymbol{\epsilon} $ \tcp*{convex combination of current and old latent vectors with noise}
  }
  $\D \leftarrow E_d(G(\Z_{\text{mix}})) $
  
    $\mathcal{L}_{\text{REM}} \leftarrow -  \frac{1}{b} \sum_{i=1}^b \log(1 -  \disc^{(i)}))$ \tcp*{compute adversarial loss}
    $\theta_E \leftarrow \theta_E - \nabla_{\theta_E}\mathcal{L}_{\text{REM}}$ \\
    $\theta_G \leftarrow \theta_G + \nabla_{\theta_G}\mathcal{L}_{\text{REM}}$ \tcp*{gradient ascent on discriminator loss}
    $\Z_{\text{old}} \leftarrow \Z $ \tcp*{keep current vectors for next iteration}
 }
 \caption{Training procedure}
 \label{algo:training}
\end{algorithm}

\subsubsection{Loss functions}

\paragraph{Wake}
In the Wake condition, we minimize the following objective function, composed of a loss for image encoding, a regularization, and a real/fake (external/internal) discriminator,
\begin{align}
 \mathcal{L}_{\text{Wake}} = \mathcal{L}_{\text{img}} +  \mathcal{L}_{\text{KL}} + \mathcal{L}_{\text{real}} \,.
\label{eqn:wake}
\end{align}

$E_z$ and $G$ learn to reconstruct the mini-batch of images $\X = \{\x^{(1)} , \x^{(2)}, ..., \x^{(b)} \}$ similarly to autoencoders \citep{bengio_representation_2013} by minimizing the image reconstruction loss $\mathcal{L}_\text{img}$ defined by
\begin{align}
  \mathcal{L}_{\text{img}} &=\frac{1}{b} \sum_{i=1}^b \| \x^{(i)} - G(E_z(\x^{(i)}))\|^2 \;,
  \label{eqn:L_img}
\end{align}
where $b$ denotes the size of the mini-batch.
We store the latent vectors $\Z = E_z(\X)$ corresponding to the current mini-batch for usage during the NREM and REM phases. 

We additionally impose a Kullback-Leibler divergence loss on the encoder $E_z$.
This acts as a regularizer and encourages latent activities to be Gaussian with zero mean and unit variance:
\begin{align}
  \mathcal{L}_{\text{KL}} &= \text{D}_{\text{KL}} ( q(\Z | \X) || p(\Z) ) \;,
\end{align}   
where $q(\Z | \X)  \sim \mathcal{N}(\boldsymbol{\mu}, \boldsymbol{\sigma}^2) $ is a distribution over the latent variables $\Z$, parametrized by mean $\boldsymbol{\mu}$ and standard deviation $\boldsymbol{\sigma}$, and $p(\Z) \sim \mathcal{N}(0,1)$ is the prior distribution over latent variables.
$E_z$ is trained to minimize the following loss:
\begin{align}
  \mathcal{L}_{\text{KL}} = \frac{1}{2 n_z} \sum_{j=1}^{n_z} \left( {\mu_j^{(\Z)}}^2 + {\sigma_j^{(\Z)}}^2 -1 - \log{ (\sigma_j^{(\Z)}}^2 ) \right) \;,
\end{align}
where $n_z$ denotes the dimension of the latent space and where $\mu_j^{(\Z)}$ and $\sigma_j^{(\Z)}$ represent the $j$\textsuperscript{th} elements of respectively the empirical mean $\boldsymbol{\mu}^{(\Z)}$ and empirical standard deviation $\boldsymbol{\sigma}^{(\Z)}$ of the set of latent vectors $E_z(\X) = \Z$.

As part of the adversarial game, $E_d$ is trained to classify the mini-batch of images as real.
This corresponds to minimizing the loss defined as sum across the mini-batch size $b$,
\begin{align}
  \mathcal{L}_{\text{real}} &=  \mathcal{L}_{\text{GAN}} (E_d(\X), 1)
                              =  - \frac{1}{b} \sum_{i=1}^b \log(E_d(\x^{(i)})) \;.
\end{align}
Note that, in principle, $\mathcal{L}_{\text{GAN}}$ can be any GAN-specific loss function \citep{gui_review_2020}.
Here we choose the binary cross-entropy loss.

\paragraph{NREM sleep}
Each Wake phase is followed by a NREM phase. During this phase we make use of the mini-batch of latent vectors $\z$ stored during the Wake phase.
Starting from a mini-batch of latent vectors, we generate images $G(z)$.
Each obtained image of $G(\z)$ is multiplied by a binary occlusion mask $\boldsymbol{\omega}$ of the same dimension.
This mask is generated by randomly picking two occlusion parameters, occlusion intensity and square size (for details see Sec.~\ref{sec:methods-image-occlusion}).
The encoder $E_z$ learns to reconstruct the latent vectors $\z$ by minimizing the following reconstruction loss:
\begin{align}
 \mathcal{L_{\text{NREM}}} = \frac{1}{b} \sum_{i=1}^b  \|\z^{(i)} - E_z\left( G(\z^{(i)}) \odot \boldsymbol{\omega} \right)\|^2 \;,
\label{eqn:nrem} 
\end{align}
where $\odot$ denotes the element-wise product.

\paragraph{REM sleep}
In REM, each latent vector from the mini-batch considered during Wake is combined with the latent vector from the previous mini-batch, the whole being convex combined with a mini-batch of noise vectors $\boldsymbol{\epsilon} \sim \mathcal{N}(0,I)$, where $I$ is the identity matrix, leading to a mini-batch of latent vectors $\Z_{\text{mix}} = \lambda' (\lambda \Z + (1 - \lambda) \Z_{\text{old}}) + (1-\lambda') \boldsymbol{\epsilon}$.
Here,  $\lambda = 0.5$ and $\lambda' = 0.5$, where $\Z_{\text{old}}$ is the previous mini-batch of latent activities.
This batch of latent vectors is passed through $G$ to generate the associated images $G(\Z_{\text{mix}})$.
In this phase, the loss function encourages $E_d$ to classify $G(\Z_{\text{mix}})$ as fake, while adversarially pushing $G$ to generate images which are less likely to be classified as fake by the minimax objective
\begin{align}
  \min_{E_d} \max_{G} \mathcal{L}_{\text{REM}} \;,
\end{align}
where
\begin{align}
  \mathcal{L}_{\text{REM}} &= \mathcal{L}_{\text{GAN}} (E_d(G(\Z_\lambda)), 0) 
                             = - \frac{1}{b}\sum_{i=1}^{b} \log(1 -  E_d(G(\mathbf{\z_{\lambda}}^{(i)})) \;.
 \label{eq:minmax}
\end{align}
In our model, the adversarial process is simply described by a full backpropagation of error through $E_d$ and $G$ with a sign switch of weight changes in $G$.

In summary, each Wake-NREM-REM cycle consists of: 1) reconstructing a mini-batch $\x$ of images during Wake, 2) reconstructing a mini-batch of latent activities $\Z=E_z(\X)$ during NREM with perturbation of $G(z)$, and 3) replaying $\Z$ convex combined with $\Z_{\text{old}}$ and noise from the $(n-1)$-th cycle.
In PAD training, all losses are weighted equally and we did not use a schedule for $\mathcal{L}_{\text{KL}}$, as opposed to standard Variational Autoencoder (VAE) training \citep{kingma2013autoencoding}.
One training epoch is defined by the number of mini-batches necessary to cover the whole dataset.
The evolution of losses with training epochs is shown in Fig.~\ref{fig:supp_losses_cifar10} and Fig.~\ref{fig:supp_losses_svhn}. The whole training procedure is summarized in the pseudo-code implemented in Algorithm~\ref{algo:training}.

\subsubsection{Image occlusion}
\label{sec:methods-image-occlusion}

\begin{figure}[!h]
  \centering
  \includegraphics[width=0.5\textwidth]{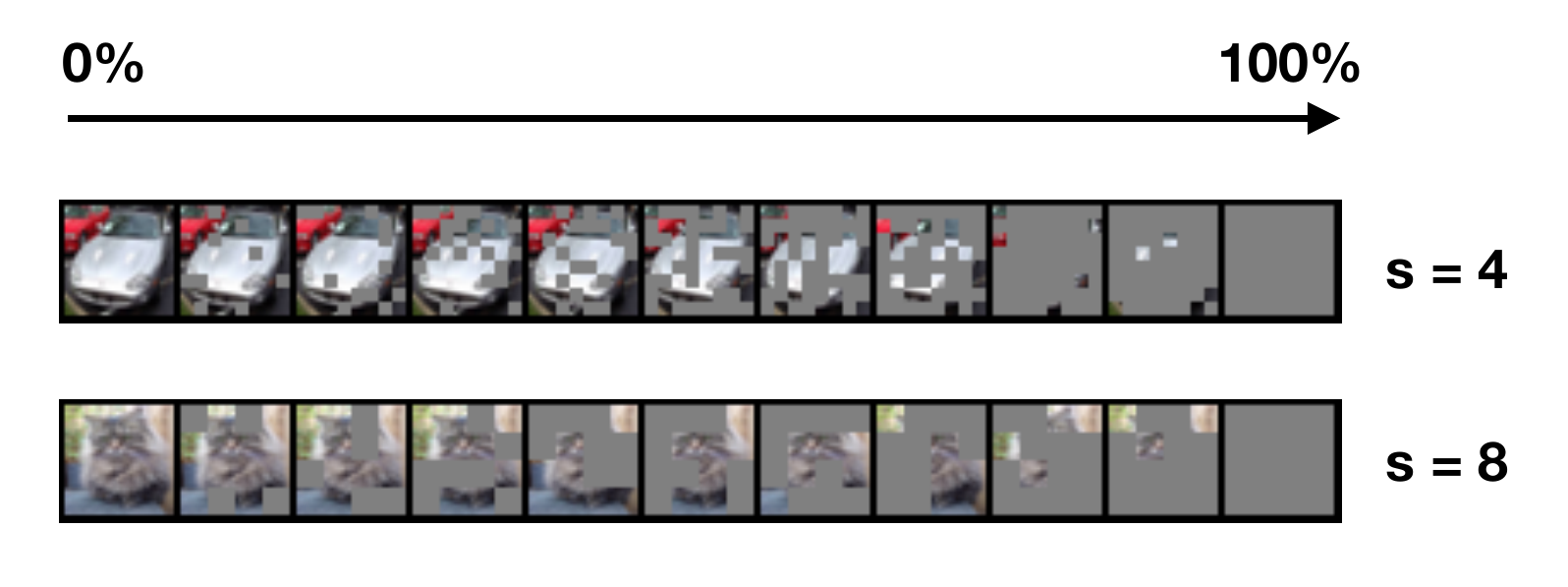}
  \caption{Varying size and intensity of occlusions on example images from CIFAR-10.
      Image occlusions vary along 2 parameters: occlusion intensity, defined by the probability to apply a grey square at a given position, and square size (s).
  }\label{fig:methods_occlusion}
\end{figure}

Following previous work \citep{zeiler_visualizing_2013}, grey squares of various sizes are applied along the image with a certain probability (Fig.~\ref{fig:methods_occlusion}).
For each mini-batch, a probability and square size were randomly picked between $0$ and $1$, and $1-8$ respectively.
We divide the image into patches of the given size and we replace each patch with a constant value (here, 0) according to the defined probability.

\subsection{Evaluation}

\subsubsection{Training of linear read-out}

A linear classifier is trained on top of latent features $\Z = E_z(\X)$, with $\Z \in \mathbb{R}^{N \times 256}$, where $N$ is the number of training dataset images.
A latent feature $\z \in \mathbb{R}^{256}$ is projected via a weight matrix $W \in \mathbb{R}^{10 \times 256}$ to the label neurons to obtain the vector $\y = W\z$.

This weight matrix is trained in a supervised fashion by using a multi-class cross-entropy loss.
For a feature $\z$ labelled with a target class $t \in \{0, 1,..,9\}$, the per-sample classification loss is given by
\begin{align}
  \mathcal{L}^C(\z,t;W) &= -\log p_{W} (Y=t | \z) \;.
 \label{eq:classifier_loss}
\end{align}
Here, $p_{\textbf{W}}$ is the conditional probability of the classifier defined by the linear projection and the softmax function 
\begin{align}
 p_{W} (Y=t | \z) &= \frac{e^{y_t}}{\sum_{i=0}^{9} e^{y_i}} \;.
 \label{eq:softmax}
\end{align}
The classifier is trained by mini-batch ($b=64$) stochastic gradient descent on the loss $\mathcal{L}^C$ with a learning rate $\eta = 0.2$ for $20$ epochs, using the whole training dataset.

\subsubsection{Linear separability}
Following previous work \citep{hjelm_learning_2019}, we define linear separability as the classification accuracy of the trained classifier on inferred latent activities $E_z(\X_{\text{test}})$ from a separate test dataset $\X_{\text{test}}$.
Given a latent feature $\z$, class prediction is made by picking the index of the maximal activity in the vector $\y$.
We ran several simulations for 4 different initial parameters of $E$ and $G$ and report the average test accuracy and standard error of the mean over trials.
To evaluate performance on occluded data, we applied random square occlusion masks on each sample from $\X_{\text{test}}$ for a fixed probability of occlusion and square size.
We report only results for occulusions of size $4$, after observing similar results with other square sizes.

\subsubsection{PCA visualization}
To visualize the 256-dimensional latent representation $E_z(\x)$ of the trained model we used the Principal Component Analysis reduction algorithm \citep{jolliffe_principal_2016}.
We project the latent representations to the first two principle components.

\subsubsection{Latent-space organization metrics}
Intra-class distance is computed by randomly picking $1,000$ pairs of images of the same class, projecting them to the encoder latent space $\z$ and computing their Euclidian distance.
This process is repeated over the $10$ classes in order to obtain the average over $10$ classes.
Similarly, inter-class distance is computed by randomly picking $10,000$ pairs of images of different classes, projecting them to the encoder latent space $\z$ and computing their Euclidian distance.
The ratio of intra- and inter-class distance is obtained by dividing the mean intra-class distance by the mean inter-class distance.
Clean-occluded distance is computed by randomly picking $10,000$ pairs of non-occluded/occluded images, projecting them to the encoder latent space and computing their Euclidian distance.
The ratio of clean-occluded and inter-class distance is obtained by dividing the clean-occluded distance by the mean inter-class distance.
We performed this analysis for several different trained networks with different initial conditions and report the mean ratios and standard error of the mean over trials.

\subsubsection{Fr\'echet inception distance}
Following \citet{heusel_gans_2018}, Fr\'echet inception distance (FID) is computed by comparing the statistics of generated (NREM or REM) samples to real images from the training dataset projected through an Inception-v3 network pre-trained on ImageNet
\begin{align}
  \text{FID} = \| \mu_{\text{real}} - \mu_{\text{gen}} \|^2 + \text{Tr}(\Sigma_{\text{real}} + \Sigma_{\text{gen}} - 2 (\Sigma_{\text{real}} \Sigma_{\text{gen}})^{1/2})
\end{align}
where $\mu$ and $\Sigma$ represent the empirical mean and covariance of the $2048$-dimensional activations of the Inception v3 pool3 layer for $10,000$ pairs of data samples and generated images.
Results represent mean FID and standard error of the mean FID over $4$ different trained networks with different initializations.

\subsubsection{Modifications specific to pathological models}

To evaluate the differential effects of each phase, we removed NREM and/or REM phases from training (Fig.~\ref{fig:results_linear_REM}, \ref{fig:results_linear_NREM}, \ref{fig:results_PCA}).
For instance, for the condition w/o NREM, the network is never trained with NREM.

A few adjustments were empirically observed to be necessary in order to obtain a fair comparison between each condition.
When removing the REM phase during training, we observed a decrease of linear separaribility after some ($>25$) epochs.
We suspect that this decrease is a result of overfitting due to unconstrained autoencoding objective of $E$ and $G$.
Models trained without REM hence would not provide a good baseline to reveal the effect of adversarial dreaming on linear separability.
For models without the REM phase, we hence added a vector of Gaussian noise $ \boldsymbol{\epsilon} \sim \mathcal{N}(0,0.5 \cdot I)$ to the encoded activities $E_z(\X)$ of dimension $n_z$ before feeding them to the generator.
Thus, Eq.~\ref{eqn:L_img} becomes: 
\begin{align}
  \mathcal{L}_{\text{img}} &= \frac{1}{b} \sum_{i=1}^b \| \x^{(i)} - G\left( E_z(\x^{(i)}) + \boldsymbol{\epsilon} \right)\|^2 \;,
\end{align}
which stabilizes linear separability of latent activities around its maximal value for both CIFAR-10 and SVHN datasets until the end of training.

Furthermore, we observed that the NREM phase alters linear performance in the absence of REM (w/o REM condition).
To overcome this issue, we reduced the effect of NREM by scaling down its loss with a factor of $0.5$.
This enabled to benefit from NREM (recognition under image occlusion) without altering linear separability on full images.

\section{Acknowledgements}

This work has received funding from the European Union 7th Framework Programme under grant agreement 604102 (HBP), the Horizon 2020 Framework Programme under grant agreements 720270, 785907 and 945539 (HBP), the Swiss National Science Foundation (SNSF, Sinergia grant CRSII5-180316), the Interfaculty Research Cooperation (IRC) `Decoding Sleep' of the University of Bern, and the Manfred St\"ark Foundation.
The authors thank the IRC collaborators Paolo Favaro for inspiring discussions on related methods in AI and deep learning, and Antoine Adamantidis and Christoph Nissen for helpful discussions on REM/NREM sleep phenomena in mice and humans.  

\bibliographystyle{apalike}
\bibliography{manuscript.bib}

\begin{thebibliography}{}

\bibitem[Adamantidis et~al., 2019]{Adamantidis2019}
Adamantidis, A.~R., {Gutierrez Herrera}, C., and Gent, T.~C. (2019).
\newblock {Oscillating circuitries in the sleeping brain}.
\newblock {\em Nature Reviews Neuroscience}, 20(12):746--762.

\bibitem[Alemi et~al., 2018]{alemi2018fixing}
Alemi, A.~A., Poole, B., Fischer, I., Dillon, J.~V., Saurous, R.~A., and
  Murphy, K. (2018).
\newblock Fixing a broken elbo.

\bibitem[Aru et~al., 2020]{aru_apical_2020}
Aru, J., Siclari, F., Phillips, W.~A., and Storm, J.~F. (2020).
\newblock Apical drive---{A} cellular mechanism of dreaming?
\newblock {\em Neuroscience \& Biobehavioral Reviews}, 119:440--455.

\bibitem[Aru et~al., 2019]{Aru2019}
Aru, J., Suzuki, M., Rutiku, R., Larkum, M.~E., and Bachmann, T. (2019).
\newblock {Coupling the State and Contents of Consciousness}.
\newblock {\em Frontiers in Systems Neuroscience}, 13(August):1--9.

\bibitem[Avitan et~al., 2021]{avitan2021spontaneous}
Avitan, L., Pujic, Z., M{\"o}lter, J., Zhu, S., Sun, B., and Goodhill, G.~J.
  (2021).
\newblock Spontaneous and evoked activity patterns diverge over development.
\newblock {\em Elife}, 10:e61942.

\bibitem[Baird et~al., 2019]{Baird2019}
Baird, B., Mota-Rolim, S.~A., and Dresler, M. (2019).
\newblock {The cognitive neuroscience of lucid dreaming}.
\newblock {\em Neuroscience {\&} Biobehavioral Reviews}, 100(May
  2018):305--323.

\bibitem[Bang et~al., 2020]{bang_discriminator_2020}
Bang, D., Kang, S., and Shim, H. (2020).
\newblock Discriminator feature-based inference by recycling the discriminator
  of gans.
\newblock {\em International Journal of Computer Vision},
  128(10-11):2436--2458.

\bibitem[Beckham et~al., 2019]{beckham_adversarial_nodate}
Beckham, C., Honari, S., Verma, V., Lamb, A.~M., Ghadiri, F., Hjelm, R.~D.,
  Bengio, Y., and Pal, C. (2019).
\newblock On {Adversarial} {Mixup} {Resynthesis}.
\newblock page~12.

\bibitem[Bengio et~al., 2013]{bengio_representation_2013}
Bengio, Y., Courville, A., and Vincent, P. (2013).
\newblock Representation {Learning}: {A} {Review} and {New} {Perspectives}.
\newblock {\em IEEE Transactions on Pattern Analysis and Machine Intelligence},
  35(8):1798--1828.

\bibitem[Benjamin and Kording, 2021]{benjamin_learning_2021}
Benjamin, A.~S. and Kording, K.~P. (2021).
\newblock Learning to infer in recurrent biological networks.
\newblock {\em arXiv:2006.10811 [cs, q-bio, stat]}.
\newblock arXiv: 2006.10811.

\bibitem[Bergelson and Swingley, 2012]{bergelson_at_2012}
Bergelson, E. and Swingley, D. (2012).
\newblock At 6-9 months, human infants know the meanings of many common nouns.
\newblock {\em Proceedings of the National Academy of Sciences},
  109(9):3253--3258.

\bibitem[Berkes et~al., 2011]{berkes2011spontaneous}
Berkes, P., Orb{\'a}n, G., Lengyel, M., and Fiser, J. (2011).
\newblock Spontaneous cortical activity reveals hallmarks of an optimal
  internal model of the environment.
\newblock {\em Science}, 331(6013):83--87.

\bibitem[Berthelot et~al., 2018]{berthelot_understanding_2018}
Berthelot, D., Raffel, C., Roy, A., and Goodfellow, I. (2018).
\newblock Understanding and {Improving} {Interpolation} in {Autoencoders} via
  an {Adversarial} {Regularizer}.
\newblock {\em arXiv:1807.07543 [cs, stat]}.
\newblock arXiv: 1807.07543.

\bibitem[Bornschein and Bengio, 2015]{bornschein2015reweighted}
Bornschein, J. and Bengio, Y. (2015).
\newblock Reweighted wake-sleep.

\bibitem[Boyce et~al., 2016]{boyce_causal_2016}
Boyce, R., Glasgow, S., Williams, S., and Adamantidis, A. (2016).
\newblock Causal evidence for the role of rem sleep theta rhythm in contextual
  memory consolidation.
\newblock {\em Science}, 352:812 -- 816.

\bibitem[Boyce et~al., 2017]{boyce_rem_2017}
Boyce, R., Williams, S., and Adamantidis, A. (2017).
\newblock {REM} sleep and memory.
\newblock {\em Current Opinion in Neurobiology}, 44:167--177.

\bibitem[Brendel and Bethge, 2019]{brendel2019approximating}
Brendel, W. and Bethge, M. (2019).
\newblock Approximating cnns with bag-of-local-features models works
  surprisingly well on imagenet.
\newblock {\em arXiv preprint arXiv:1904.00760}.

\bibitem[Brock et~al., 2017]{brock_neural_2017}
Brock, A., Lim, T., Ritchie, J.~M., and Weston, N. (2017).
\newblock Neural {Photo} {Editing} with {Introspective} {Adversarial}
  {Networks}.
\newblock {\em arXiv:1609.07093 [cs, stat]}.
\newblock arXiv: 1609.07093.

\bibitem[Buzs{\'a}ki, 2002]{buzsaki_theta_2002}
Buzs{\'a}ki, G. (2002).
\newblock Theta {Oscillations} in the {Hippocampus}.
\newblock {\em Neuron}, 33(3):325--340.

\bibitem[Cai et~al., 2009]{cai_rem_2009}
Cai, D.~J., Mednick, S.~A., Harrison, E.~M., Kanady, J.~C., and Mednick, S.~C.
  (2009).
\newblock {REM}, not incubation, improves creativity by priming associative
  networks.
\newblock {\em Proceedings of the National Academy of Sciences},
  106(25):10130--10134.

\bibitem[Chen et~al., 2020]{chen_simple_2020}
Chen, T., Kornblith, S., Norouzi, M., and Hinton, G. (2020).
\newblock A {Simple} {Framework} for {Contrastive} {Learning} of {Visual}
  {Representations}.
\newblock {\em arXiv:2002.05709 [cs, stat]}.
\newblock arXiv: 2002.05709.

\bibitem[Cohrs, 2008]{cohrs_sleep_2008}
Cohrs, S. (2008).
\newblock Sleep {Disturbances} in {Patients} with {Schizophrenia}.
\newblock {\em CNS Drugs}, 22(11):939--962.

\bibitem[Crick and Mitchison, 1983]{Crick1983}
Crick, F. and Mitchison, G. (1983).
\newblock {The function of dream sleep}.
\newblock {\em Nature}, 304(5922):111--114.

\bibitem[Dayan et~al., 1995]{dayan_helmholtz_1995}
Dayan, P., Hinton, G.~E., Neal, R.~M., and Zemel, R.~S. (1995).
\newblock {The Helmholtz Machine}.
\newblock {\em Neural Computation}, 7(5):889--904.

\bibitem[DiCarlo et~al., 2012]{dicarlo_how_2012}
DiCarlo, J.~J., Zoccolan, D., and Rust, N.~C. (2012).
\newblock How {Does} the {Brain} {Solve} {Visual} {Object} {Recognition}?
\newblock {\em Neuron}, 73(3):415--434.

\bibitem[Diekelmann and Born, 2010]{diekelmann_memory_2010}
Diekelmann, S. and Born, J. (2010).
\newblock The memory function of sleep.
\newblock {\em Nature Reviews Neuroscience}, 11(2):114--126.

\bibitem[Donahue et~al., 2016]{donahue_adversarial_2016}
Donahue, J., Kr{\"a}henb{\"u}hl, P., and Darrell, T. (2016).
\newblock Adversarial {Feature} {Learning}.
\newblock {\em arXiv:1605.09782 [cs, stat]}.
\newblock arXiv: 1605.09782.

\bibitem[Dresler et~al., 2012]{Dresler2012}
Dresler, M., Wehrle, R., Spoormaker, V.~I., Koch, S.~P., Holsboer, F., Steiger,
  A., Obrig, H., S{\"{a}}mann, P.~G., and Czisch, M. (2012).
\newblock {Neural correlates of dream lucidity obtained from contrasting lucid
  versus non-lucid REM sleep: A combined EEG/fMRI case study}.
\newblock {\em Sleep}, 35(7):1017--1020.

\bibitem[Dudai et~al., 2015]{dudai_consolidation_2015}
Dudai, Y., Karni, A., and Born, J. (2015).
\newblock The {Consolidation} and {Transformation} of {Memory}.
\newblock {\em Neuron}, 88(1):20--32.

\bibitem[Dumoulin et~al., 2017]{dumoulin_adversarially_2017}
Dumoulin, V., Belghazi, I., Poole, B., Mastropietro, O., Lamb, A., Arjovsky,
  M., and Courville, A. (2017).
\newblock Adversarially {Learned} {Inference}.
\newblock {\em arXiv:1606.00704 [cs, stat]}.
\newblock arXiv: 1606.00704.

\bibitem[Fiete et~al., 2007]{fiete_birdsong_2007}
Fiete, I.~R., Fee, M.~S., and Seung, H.~S. (2007).
\newblock Model of birdsong learning based on gradient estimation by dynamic
  perturbation of neural conductances.
\newblock {\em Journal of Neurophysiology}, 98(4):2038--2057.
\newblock PMID: 17652414.

\bibitem[Fosse et~al., 2003]{fosse_dreaming_2003}
Fosse, M.~J., Fosse, R., Hobson, J.~A., and Stickgold, R.~J. (2003).
\newblock Dreaming and episodic memory: A functional dissociation?
\newblock {\em Journal of Cognitive Neuroscience}, 15(1):1--9.

\bibitem[Foulkes, 1999]{foulkes_children_1999}
Foulkes, D. (1999).
\newblock {\em Children's dreaming and the development of consciousness}.
\newblock Harvard University Press.

\bibitem[Gershman, 2019]{gershman_generative_2019}
Gershman, S.~J. (2019).
\newblock The {Generative} {Adversarial} {Brain}.
\newblock {\em Frontiers in Artificial Intelligence}, 2.

\bibitem[Gidaris et~al., 2018]{gidaris_unsupervised_2018}
Gidaris, S., Singh, P., and Komodakis, N. (2018).
\newblock Unsupervised {Representation} {Learning} by {Predicting} {Image}
  {Rotations}.
\newblock {\em arXiv:1803.07728 [cs]}.
\newblock arXiv: 1803.07728.

\bibitem[Gilbert and Li, 2013]{gilbert_top-down_2013}
Gilbert, C.~D. and Li, W. (2013).
\newblock Top-down influences on visual processing.
\newblock {\em Nature Reviews Neuroscience}, 14(5):350--363.

\bibitem[Giuditta et~al., 1995]{giuditta1998}
Giuditta, A., Ambrosini, M.~V., Montagnese, P., Mandile, P., Cotugno, M.,
  Zucconi, G.~G., and Vescia, S. (1995).
\newblock The sequential hypothesis of the function of sleep.
\newblock {\em Behavioural Brain Research}, 69(1):157 -- 166.
\newblock The Function of Sleep.

\bibitem[Goodfellow, 2016]{goodfellow_nips_2016}
Goodfellow, I. (2016).
\newblock {NIPS} 2016 {Tutorial}: {Generative} {Adversarial} {Networks}.
\newblock {\em arXiv:1701.00160 [cs]}.
\newblock arXiv: 1701.00160.

\bibitem[Goodfellow et~al., 2014]{goodfellow2014generative}
Goodfellow, I.~J., Pouget-Abadie, J., Mirza, M., Xu, B., Warde-Farley, D.,
  Ozair, S., Courville, A., and Bengio, Y. (2014).
\newblock Generative adversarial networks.

\bibitem[Grill-Spector et~al., 2001]{grill-spector_2001_the}
Grill-Spector, K., Kourtzi, Z., and Kanwisher, N. (2001).
\newblock The lateral occipital complex and its role in object recognition.
\newblock {\em Vision Research}, 41(10):1409 -- 1422.

\bibitem[Guerguiev et~al., 2017]{guerguiev_towards_2017}
Guerguiev, J., Lillicrap, T.~P., and Richards, B.~A. (2017).
\newblock Towards deep learning with segregated dendrites.
\newblock {\em eLife}, 6:e22901.

\bibitem[Gui et~al., 2020]{gui_review_2020}
Gui, J., Sun, Z., Wen, Y., Tao, D., and Ye, J. (2020).
\newblock A {Review} on {Generative} {Adversarial} {Networks}: {Algorithms},
  {Theory}, and {Applications}.
\newblock {\em arXiv:2001.06937 [cs, stat]}.
\newblock arXiv: 2001.06937.

\bibitem[Guo et~al., 2019]{guo_multimodal_2019}
Guo, W., Wang, J., and Wang, S. (2019).
\newblock Deep multimodal representation learning: A survey.
\newblock {\em IEEE Access}, 7:63373--63394.

\bibitem[Ha and Schmidhuber, 2018]{ha2018world}
Ha, D. and Schmidhuber, J. (2018).
\newblock World models.
\newblock {\em arXiv preprint arXiv:1803.10122}.

\bibitem[Haider et~al., 2021]{haider2021latent}
Haider, P., Ellenberger, B., Kriener, L., Jordan, J., Senn, W., and Petrovici,
  M. (2021).
\newblock Latent equilibrium: Arbitrarily fast computation with arbitrarily
  slow neurons.
\newblock {\em Advances in Neural Information Processing Systems}, 34.

\bibitem[Heusel et~al., 2018]{heusel_gans_2018}
Heusel, M., Ramsauer, H., Unterthiner, T., Nessler, B., and Hochreiter, S.
  (2018).
\newblock {GANs} {Trained} by a {Two} {Time}-{Scale} {Update} {Rule} {Converge}
  to a {Local} {Nash} {Equilibrium}.
\newblock {\em arXiv:1706.08500 [cs, stat]}.
\newblock arXiv: 1706.08500.

\bibitem[Hinton et~al., 1995]{hinton_wake-sleep_1995}
Hinton, G., Dayan, P., Frey, B., and Neal, R. (1995).
\newblock The "wake-sleep" algorithm for unsupervised neural networks.
\newblock {\em Science}, 268(5214):1158--1161.

\bibitem[Hjelm et~al., 2019]{hjelm_learning_2019}
Hjelm, R.~D., Fedorov, A., Lavoie-Marchildon, S., Grewal, K., Bachman, P.,
  Trischler, A., and Bengio, Y. (2019).
\newblock Learning deep representations by mutual information estimation and
  maximization.
\newblock {\em arXiv:1808.06670 [cs, stat]}.
\newblock arXiv: 1808.06670.

\bibitem[Hobson, 2009]{hobson_rem_2009}
Hobson, J.~A. (2009).
\newblock {REM} sleep and dreaming: towards a theory of protoconsciousness.
\newblock {\em Nature Reviews Neuroscience}, 10(11):803--813.

\bibitem[Hobson et~al., 2014]{hobson_virtual_2014}
Hobson, J.~A., Hong, C. C.-H., and Friston, K.~J. (2014).
\newblock Virtual reality and consciousness inference in dreaming.
\newblock {\em Frontiers in Psychology}, 5.

\bibitem[Hobson et~al., 2000]{hobson_dreaming_2000}
Hobson, J.~A., Pace-Schott, E.~F., and Stickgold, R. (2000).
\newblock Dreaming and the brain: {Toward} a cognitive neuroscience of
  conscious states.
\newblock {\em Behavioral and Brain Sciences}, 23(6):793--842.

\bibitem[Hoel, 2021]{hoel_overfitted_2021}
Hoel, E. (2021).
\newblock The overfitted brain: {Dreams} evolved to assist generalization.
\newblock {\em Patterns}, 2(5):100244.

\bibitem[Huang et~al., 2018]{huang_introVAE_2018}
Huang, H., Li, Z., He, R., Sun, Z., and Tan, T. (2018).
\newblock Introvae: Introspective variational autoencoders for photographic
  image synthesis.

\bibitem[Hung et~al., 2005]{hung_fast_2005}
Hung, C.~P., Kreiman, G., Poggio, T., and DiCarlo, J.~J. (2005).
\newblock Fast {Readout} of {Object} {Identity} from {Macaque} {Inferior}
  {Temporal} {Cortex}.
\newblock {\em Science}, 310(5749):863--866.

\bibitem[Ioffe and Szegedy, 2015]{loffe_batch_2018}
Ioffe, S. and Szegedy, C. (2015).
\newblock Batch normalization: Accelerating deep network training by reducing
  internal covariate shift.
\newblock {\em CoRR}, abs/1502.03167.

\bibitem[Ji and Wilson, 2007]{ji_coordinated_2007}
Ji, D. and Wilson, M.~A. (2007).
\newblock Coordinated memory replay in the visual cortex and hippocampus during
  sleep.
\newblock {\em Nature Neuroscience}, 10(1):100--107.

\bibitem[Jolliffe and Cadima, 2016]{jolliffe_principal_2016}
Jolliffe, I.~T. and Cadima, J. (2016).
\newblock Principal component analysis: a review and recent developments.
\newblock {\em Philosophical Transactions of the Royal Society A: Mathematical,
  Physical and Engineering Sciences}, 374(2065):20150202.

\bibitem[K{\'a}li and Dayan, 2004]{kali_off-line_2004}
K{\'a}li, S. and Dayan, P. (2004).
\newblock Off-line replay maintains declarative memories in a model of
  hippocampal-neocortical interactions.
\newblock {\em Nature Neuroscience}, 7(3):286--294.

\bibitem[Karras et~al., 2018]{karras_style-based_2018}
Karras, T., Laine, S., and Aila, T. (2018).
\newblock A {Style}-{Based} {Generator} {Architecture} for {Generative}
  {Adversarial} {Networks}.
\newblock {\em arXiv:1812.04948 [cs, stat]}.
\newblock arXiv: 1812.04948.

\bibitem[Keller and Mrsic-Flogel, 2018]{Keller2018}
Keller, G.~B. and Mrsic-Flogel, T.~D. (2018).
\newblock {Predictive Processing: A Canonical Cortical Computation}.
\newblock {\em Neuron}, 100(2):424--435.

\bibitem[Kingma and Ba, 2017]{kingma_adam_2017}
Kingma, D.~P. and Ba, J. (2017).
\newblock Adam: {A} {Method} for {Stochastic} {Optimization}.
\newblock {\em arXiv:1412.6980 [cs]}.
\newblock arXiv: 1412.6980.

\bibitem[Kingma and Welling, 2013]{kingma2013autoencoding}
Kingma, D.~P. and Welling, M. (2013).
\newblock Auto-encoding variational bayes.

\bibitem[Klinzing et~al., 2019]{klinzing_mechanisms_2019}
Klinzing, J.~G., Niethard, N., and Born, J. (2019).
\newblock Mechanisms of systems memory consolidation during sleep.
\newblock {\em Nature Neuroscience}, 22(10):1598--1610.

\bibitem[Korcsak-Gorzo et~al., 2021]{korcsakgorzo2021cortical}
Korcsak-Gorzo, A., M{\"u}ller, M.~G., Baumbach, A., Leng, L., Breitwieser,
  O.~J., van Albada, S.~J., Senn, W., Meier, K., Legenstein, R., and Petrovici,
  M.~A. (2021).
\newblock Cortical oscillations implement a backbone for sampling-based
  computation in spiking neural networks.

\bibitem[Krizhevsky et~al., 2013]{cifar10}
Krizhevsky, A., Nair, V., and Hinton, G. (2013).
\newblock Cifar-10 (canadian institute for advanced research).

\bibitem[LeCun et~al., 2015]{lecun_deep_2015}
LeCun, Y., Bengio, Y., and Hinton, G. (2015).
\newblock Deep learning.
\newblock {\em Nature}, 521(7553):436--444.

\bibitem[L{\'e}ger et~al., 2018]{leger_slow-wave_2018}
L{\'e}ger, D., Debellemaniere, E., Rabat, A., Bayon, V., Benchenane, K., and
  Chennaoui, M. (2018).
\newblock Slow-wave sleep: {From} the cell to the clinic.
\newblock {\em Sleep Medicine Reviews}, 41:113--132.

\bibitem[Lewis and Durrant, 2011]{lewis_overlapping_2011}
Lewis, P.~A. and Durrant, S.~J. (2011).
\newblock Overlapping memory replay during sleep builds cognitive schemata.
\newblock {\em Trends in Cognitive Sciences}, 15(8):343--351.

\bibitem[Lewis et~al., 2018]{lewis_how_2018}
Lewis, P.~A., Knoblich, G., and Poe, G. (2018).
\newblock How {Memory} {Replay} in {Sleep} {Boosts} {Creative}
  {Problem}-{Solving}.
\newblock {\em Trends in Cognitive Sciences}, 22(6):491--503.

\bibitem[Li et~al., 2017]{Li2017}
Li, W., Ma, L., Yang, G., and Gan, W.-b. (2017).
\newblock {REM sleep selectively prunes and maintains new synapses in
  development and learning}.
\newblock {\em Nature Neuroscience}, 20(3):427--437.

\bibitem[Lillicrap et~al., 2020]{lillicrap_backpropagation_2020}
Lillicrap, T.~P., Santoro, A., Marris, L., Akerman, C.~J., and Hinton, G.
  (2020).
\newblock Backpropagation and the brain.
\newblock {\em Nature Reviews Neuroscience}.

\bibitem[Lim et~al., 2015]{Lim2015}
Lim, S., McKee, J.~L., Woloszyn, L., Amit, Y., Freedman, D.~J., Sheinberg,
  D.~L., and Brunel, N. (2015).
\newblock {Inferring learning rules from distributions of firing rates}.
\newblock {\em Nature neuroscience}, 18(12):1--13.

\bibitem[Liu et~al., 2021]{liu_self-supervised_2021}
Liu, X., Zhang, F., Hou, Z., Wang, Z., Mian, L., Zhang, J., and Tang, J.
  (2021).
\newblock Self-supervised {Learning}: {Generative} or {Contrastive}.
\newblock {\em arXiv:2006.08218 [cs, stat]}.
\newblock arXiv: 2006.08218.

\bibitem[Llewellyn, 2016a]{llewellyn_crossing_2016}
Llewellyn, S. (2016a).
\newblock Crossing the invisible line: {De}-differentiation of wake, sleep and
  dreaming may engender both creative insight and psychopathology.
\newblock {\em Consciousness and Cognition}, 46:127--147.

\bibitem[Llewellyn, 2016b]{llewellyn_dream_2016}
Llewellyn, S. (2016b).
\newblock Dream to {Predict}? {REM} {Dreaming} as {Prospective} {Coding}.
\newblock {\em Frontiers in Psychology}, 6.

\bibitem[Maas et~al., 2013]{leakyrelu}
Maas, A.~L., Hannun, A.~Y., and Ng, A.~Y. (2013).
\newblock Rectifier nonlinearities improve neural network acoustic models.
\newblock In {\em in ICML Workshop on Deep Learning for Audio, Speech and
  Language Processing}.

\bibitem[Majaj et~al., 2015]{majaj_simple_2015}
Majaj, N.~J., Hong, H., Solomon, E.~A., and DiCarlo, J.~J. (2015).
\newblock Simple {Learned} {Weighted} {Sums} of {Inferior} {Temporal}
  {Neuronal} {Firing} {Rates} {Accurately} {Predict} {Human} {Core} {Object}
  {Recognition} {Performance}.
\newblock {\em Journal of Neuroscience}, 35(39):13402--13418.

\bibitem[Mamelak and Hobson, 1989]{mamelak_dream_1989}
Mamelak, A.~N. and Hobson, J.~A. (1989).
\newblock {Dream Bizarreness as the Cognitive Correlate of Altered Neuronal
  Behavior in REM Sleep}.
\newblock {\em Journal of Cognitive Neuroscience}, 1(3):201--222.

\bibitem[Marblestone et~al., 2016]{marblestone2016toward}
Marblestone, A.~H., Wayne, G., and Kording, K.~P. (2016).
\newblock Toward an integration of deep learning and neuroscience.
\newblock {\em Frontiers in computational neuroscience}, 10:94.

\bibitem[McClelland et~al., 1995]{mcclelland_why_1995}
McClelland, J.~L., McNaughton, B.~L., and O'Reilly, R.~C. (1995).
\newblock Why there are complementary learning systems in the hippocampus and
  neocortex: {Insights} from the successes and failures of connectionist models
  of learning and memory.
\newblock {\em Psychological Review}, 102(3):419--457.

\bibitem[McKay et~al., 2007]{McKay2007}
McKay, B.~E., Placzek, A.~N., and Dani, J.~A. (2007).
\newblock {Regulation of synaptic transmission and plasticity by neuronal
  nicotinic acetylcholine receptors}.
\newblock {\em Biochemical Pharmacology}, 74(8):1120--1133.

\bibitem[Miyato et~al., 2018]{miyato_spectral_2018}
Miyato, T., Kataoka, T., Koyama, M., and Yoshida, Y. (2018).
\newblock Spectral normalization for generative adversarial networks.

\bibitem[Munjal et~al., 2019]{munjal_implicit_2019}
Munjal, P., Paul, A., and Krishnan, N.~C. (2019).
\newblock Implicit discriminator in variational autoencoder.

\bibitem[Nadel and Moscovitch, 1997]{nadel_memory_1997}
Nadel, L. and Moscovitch, M. (1997).
\newblock Memory consolidation, retrograde amnesia and the hippocampal complex.
\newblock {\em Current Opinion in Neurobiology}, 7:217--227.

\bibitem[Nayebi et~al., 2020]{nayebi2020identifying}
Nayebi, A., Srivastava, S., Ganguli, S., and Yamins, D.~L. (2020).
\newblock Identifying learning rules from neural network observables.
\newblock {\em Advances in Neural Information Processing Systems},
  33:2639--2650.

\bibitem[Nelson et~al., 1983]{nelson_rem_1983}
Nelson, J.~P., McCarley, R.~W., and Hobson, J.~A. (1983).
\newblock {REM} sleep burst neurons, {PGO} waves, and eye movement information.
\newblock {\em Journal of Neurophysiology}, 50(4):784--797.

\bibitem[Netzer et~al., 2011]{svhn}
Netzer, Y., Wang, T., Coates, A., Bissacco, A., Wu, B., and Ng, A.~Y. (2011).
\newblock Reading digits in natural images with unsupervised feature learning.

\bibitem[Nir and Tononi, 2010]{nir_dreaming_2010}
Nir, Y. and Tononi, G. (2010).
\newblock Dreaming and the brain: from phenomenology to neurophysiology.
\newblock {\em Trends in Cognitive Sciences}, 14(2):88--100.

\bibitem[Norman et~al., 2005]{norman_methods_2005}
Norman, K.~A., Newman, E.~L., and Perotte, A.~J. (2005).
\newblock Methods for reducing interference in the {Complementary} {Learning}
  {Systems} model: {Oscillating} inhibition and autonomous memory rehearsal.
\newblock {\em Neural Networks}, 18(9):1212--1228.

\bibitem[O'Neill et~al., 2010]{oneill_play_2010}
O'Neill, J., Pleydell-Bouverie, B., Dupret, D., and Csicsvari, J. (2010).
\newblock Play it again: reactivation of waking experience and memory.
\newblock {\em Trends in Neurosciences}, 33(5):220--229.

\bibitem[Palmiero et~al., 2015]{palmiero_domain-specificity_2015}
Palmiero, M., Nori, R., Aloisi, V., Ferrara, M., and Piccardi, L. (2015).
\newblock Domain-{Specificity} of {Creativity}: {A} {Study} on the
  {Relationship} {Between} {Visual} {Creativity} and {Visual} {Mental}
  {Imagery}.
\newblock {\em Frontiers in Psychology}, 6.

\bibitem[Poe, 2017]{Poe2017}
Poe, G.~R. (2017).
\newblock {Sleep is for forgetting}.
\newblock {\em Journal of Neuroscience}, 37(3):464--473.

\bibitem[Pogodin et~al., 2021]{pogodin_towards_2021}
Pogodin, R., Mehta, Y., Lillicrap, T.~P., and Latham, P.~E. (2021).
\newblock Towards {Biologically} {Plausible} {Convolutional} {Networks}.
\newblock {\em arXiv:2106.13031 [cs, q-bio]}.
\newblock arXiv: 2106.13031.

\bibitem[Radford et~al., 2021]{radford2021learning}
Radford, A., Kim, J.~W., Hallacy, C., Ramesh, A., Goh, G., Agarwal, S., Sastry,
  G., Askell, A., Mishkin, P., Clark, J., Krueger, G., and Sutskever, I.
  (2021).
\newblock Learning transferable visual models from natural language
  supervision.

\bibitem[Radford et~al., 2015]{radford_unsupervised_2015}
Radford, A., Metz, L., and Chintala, S. (2015).
\newblock Unsupervised {Representation} {Learning} with {Deep} {Convolutional}
  {Generative} {Adversarial} {Networks}.
\newblock {\em arXiv:1511.06434 [cs]}.
\newblock arXiv: 1511.06434.

\bibitem[Rao and Ballard, 1999]{rao_predictive_1999}
Rao, R. P.~N. and Ballard, D.~H. (1999).
\newblock Predictive coding in the visual cortex: a functional interpretation
  of some extra-classical receptive-field effects.
\newblock {\em Nature Neuroscience}, 2(1):79--87.

\bibitem[Renn{\'o}-Costa et~al., 2019]{renno-costa_computational_2019}
Renn{\'o}-Costa, C., da~Silva, A. C.~C., Blanco, W., and Ribeiro, S. (2019).
\newblock Computational models of memory consolidation and long-term synaptic
  plasticity during sleep.
\newblock {\em Neurobiology of Learning and Memory}, 160:32--47.

\bibitem[Richards et~al., 2019]{richards_2019}
Richards, B.~A., Lillicrap, T.~P., Beaudoin, P., Bengio, Y., Bogacz, R.,
  Christensen, A., Clopath, C., Costa, R.~P., de~Berker, A., Ganguli, S.,
  Gillon, C.~J., Hafner, D., Kepecs, A., Kriegeskorte, N., Latham, P., Lindsay,
  G.~W., Miller, K.~D., Naud, R., Pack, C.~C., Poirazi, P., Roelfsema, P.,
  Sacramento, J., Saxe, A., Scellier, B., Schapiro, A.~C., Senn, W., Wayne, G.,
  Yamins, D., Zenke, F., Zylberberg, J., Therien, D., and Kording, K.~P.
  (2019).
\newblock A deep learning framework for neuroscience.
\newblock {\em Nature Neuroscience}, 22(11):1761--1770.

\bibitem[Sacramento et~al., 2018]{sacramento_dendritic_2017}
Sacramento, J.~a., Ponte~Costa, R., Bengio, Y., and Senn, W. (2018).
\newblock Dendritic cortical microcircuits approximate the backpropagation
  algorithm.
\newblock In Bengio, S., Wallach, H., Larochelle, H., Grauman, K.,
  Cesa-Bianchi, N., and Garnett, R., editors, {\em Advances in Neural
  Information Processing Systems}, volume~31. Curran Associates, Inc.

\bibitem[Schoenfeld et~al., 2022]{Schoenfeld2022}
Schoenfeld, G., Kollmorgen, S., Lewis, C., Bethge, P., Ruess, A. M.~A., Aguzzi,
  A., Mante, V., and Helmchen, F. (2022).
\newblock {Dendritic integration of sensory and reward information facilitates
  learning}.
\newblock {\em bioRxiv}, pages 1--31.

\bibitem[Schrittwieser et~al., 2020]{schrittwieser2020mastering}
Schrittwieser, J., Antonoglou, I., Hubert, T., Simonyan, K., Sifre, L.,
  Schmitt, S., Guez, A., Lockhart, E., Hassabis, D., Graepel, T., et~al.
  (2020).
\newblock Mastering atari, go, chess and shogi by planning with a learned
  model.
\newblock {\em Nature}, 588(7839):604--609.

\bibitem[Schwartz, 2003]{schwartz_are_2003}
Schwartz, S. (2003).
\newblock Are life episodes replayed during dreaming?
\newblock {\em Trends in Cognitive Sciences}, 7(8):325--327.

\bibitem[Seibt et~al., 2017]{Seibt2017}
Seibt, J., Richard, C.~J., Sigl-Gl{\"{o}}ckner, J., Takahashi, N., Kaplan,
  D.~I., Doron, G., Limoges, D.~D., Bocklisch, C., and Larkum, M.~E. (2017).
\newblock {Cortical dendritic activity correlates with spindle-rich
  oscillations during sleep in rodents}.
\newblock {\em Nature Communications}, 8(684):1--13.

\bibitem[Senn and Sacramento, 2015]{Senn2015}
Senn, W. and Sacramento, J. (2015).
\newblock {Backward reasoning the formation rules}.
\newblock {\em Nature Neuroscience}, 18(12):1705--1706.

\bibitem[Shorten and Khoshgoftaar, 2019]{shorten2019survey}
Shorten, C. and Khoshgoftaar, T.~M. (2019).
\newblock A survey on image data augmentation for deep learning.
\newblock {\em Journal of Big Data}, 6(1):1--48.

\bibitem[Siegel, 2009]{siegel_sleep_2009}
Siegel, J.~M. (2009).
\newblock Sleep viewed as a state of adaptive inactivity.
\newblock {\em Nature Reviews Neuroscience}, 10(10):747--753.

\bibitem[Silver et~al., 2017]{silver2017predictron}
Silver, D., Hasselt, H., Hessel, M., Schaul, T., Guez, A., Harley, T.,
  Dulac-Arnold, G., Reichert, D., Rabinowitz, N., Barreto, A., et~al. (2017).
\newblock The predictron: End-to-end learning and planning.
\newblock In {\em International Conference on Machine Learning}, pages
  3191--3199. PMLR.

\bibitem[Simons et~al., 2017]{simons_brain_2017}
Simons, J.~S., Garrison, J.~R., and Johnson, M.~K. (2017).
\newblock Brain mechanisms of reality monitoring.
\newblock {\em Trends in Cognitive Sciences}, 21(6):462--473.

\bibitem[Sj{\"{o}}str{\"{o}}m and H{\"{a}}usser, 2006]{Sjostrom2006a}
Sj{\"{o}}str{\"{o}}m, P.~J. and H{\"{a}}usser, M. (2006).
\newblock {A Cooperative Switch Determines the Sign of Synaptic Plasticity in
  Distal Dendrites of Neocortical Pyramidal Neurons}.
\newblock {\em Neuron}, 51(2):227--238.

\bibitem[Span{\`o} et~al., 2020]{spano_dreaming_2020}
Span{\`o}, G., Pizzamiglio, G., McCormick, C., Clark, I.~A., De~Felice, S.,
  Miller, T.~D., Edgin, J.~O., Rosenthal, C.~R., and Maguire, E.~A. (2020).
\newblock Dreaming with hippocampal damage.
\newblock {\em eLife}, 9.

\bibitem[Subramaniam et~al., 2012]{Subramaniam2012}
Subramaniam, K., Luks, T.~L., Fisher, M., Simpson, G.~V., Nagarajan, S., and
  Vinogradov, S. (2012).
\newblock {Computerized Cognitive Training Restores Neural Activity within the
  Reality Monitoring Network in Schizophrenia}.
\newblock {\em Neuron}, 73(4):842--853.

\bibitem[Takahashi et~al., 2020]{Takahashi2020}
Takahashi, N., Ebner, C., Sigl-Gl{\"{o}}ckner, J., Moberg, S., Nierwetberg, S.,
  and Larkum, M.~E. (2020).
\newblock {Active dendritic currents gate descending cortical outputs in
  perception}.
\newblock {\em Nature Neuroscience}, 23(10):1277--1285.

\bibitem[Tang et~al., 2010]{tang_memory_2010}
Tang, H., Li, H., and Yan, R. (2010).
\newblock Memory {Dynamics} in {Attractor} {Networks} with {Saliency}
  {Weights}.
\newblock {\em Neural Computation}, 22(7):1899--1926.

\bibitem[Tenenbaum et~al., 2011]{tenenbaum_how_2011}
Tenenbaum, J.~B., Kemp, C., Griffiths, T.~L., and Goodman, N.~D. (2011).
\newblock How to {Grow} a {Mind}: {Statistics}, {Structure}, and {Abstraction}.
\newblock {\em Science}, 331(6022):1279--1285.

\bibitem[Tononi and Cirelli, 2014]{tononi_sleep_2014}
Tononi, G. and Cirelli, C. (2014).
\newblock Sleep and the {Price} of {Plasticity}: {From} {Synaptic} and
  {Cellular} {Homeostasis} to {Memory} {Consolidation} and {Integration}.
\newblock {\em Neuron}, 81(1):12--34.

\bibitem[Tononi and Cirelli, 2020]{Tononi2020}
Tononi, G. and Cirelli, C. (2020).
\newblock {Sleep and synaptic down-selection}.
\newblock {\em European Journal of Neuroscience}, 51(1):413--421.

\bibitem[Tschannen et~al., 2020]{tschannen2020mutual}
Tschannen, M., Djolonga, J., Rubenstein, P.~K., Gelly, S., and Lucic, M.
  (2020).
\newblock On mutual information maximization for representation learning.

\bibitem[Ulyanov et~al., 2017]{ulyanov_it_2017}
Ulyanov, D., Vedaldi, A., and Lempitsky, V. (2017).
\newblock It {Takes} ({Only}) {Two}: {Adversarial} {Generator}-{Encoder}
  {Networks}.
\newblock {\em arXiv:1704.02304 [cs, stat]}.
\newblock arXiv: 1704.02304.

\bibitem[Urbanczik and Senn, 2014]{Urbanczik2014}
Urbanczik, R. and Senn, W. (2014).
\newblock {Learning by the Dendritic Prediction of Somatic Spiking}.
\newblock {\em Neuron}, 81(3):521--528.

\bibitem[van~de Ven et~al., 2020]{van_de_ven_brain-inspired_2020}
van~de Ven, G.~M., Siegelmann, H.~T., and Tolias, A.~S. (2020).
\newblock Brain-inspired replay for continual learning with artificial neural
  networks.
\newblock {\em Nature Communications}, 11(1).

\bibitem[Voigts and Harnett, 2020]{Voigts2020}
Voigts, J. and Harnett, M.~T. (2020).
\newblock {Somatic and Dendritic Encoding of Spatial Variables in Retrosplenial
  Cortex Differs during 2D Navigation}.
\newblock {\em Neuron}, 105(2):237--245.e4.

\bibitem[Walker, 2009]{walker_role_2009}
Walker, M.~P. (2009).
\newblock The {Role} of {Sleep} in {Cognition} and {Emotion}.
\newblock {\em Annals of the New York Academy of Sciences}, 1156(1):168--197.

\bibitem[Wamsley, 2014]{wamsley_dreaming_2014}
Wamsley, E.~J. (2014).
\newblock Dreaming and {Offline} {Memory} {Consolidation}.
\newblock {\em Current Neurology and Neuroscience Reports}, 14(3).

\bibitem[Waters et~al., 2016]{waters_what_2016}
Waters, F., Blom, J.~D., Dang-Vu, T.~T., Cheyne, A.~J., Alderson-Day, B.,
  Woodruff, P., and Collerton, D. (2016).
\newblock What {Is} the {Link} {Between} {Hallucinations}, {Dreams}, and
  {Hypnagogic}--{Hypnopompic} {Experiences}?
\newblock {\em Schizophrenia Bulletin}, 42(5):1098--1109.

\bibitem[Whittington and Bogacz, 2019]{whittington_theories_2019}
Whittington, J.~C. and Bogacz, R. (2019).
\newblock Theories of {Error} {Back}-{Propagation} in the {Brain}.
\newblock {\em Trends in Cognitive Sciences}, 23(3):235--250.

\bibitem[Wierzynski et~al., 2009]{wierzynski_2009}
Wierzynski, C.~M., Lubenov, E.~V., Gu, M., and Siapas, A.~G. (2009).
\newblock State-dependent spike-timing relationships between hippocampal and
  prefrontal circuits during sleep.
\newblock {\em Neuron}, 61(4):587--596.

\bibitem[Williams et~al., 1992]{williams_bizarreness_1992}
Williams, J., Merritt, J., Rittenhouse, C., and Hobson, J. (1992).
\newblock Bizarreness in dreams and fantasies: Implications for the
  activation-synthesis hypothesis.
\newblock {\em Consciousness and Cognition}, 1(2):172--185.

\bibitem[Winocur et~al., 2010]{winocur_memory_2010}
Winocur, G., Moscovitch, M., and Bontempi, B. (2010).
\newblock Memory formation and long-term retention in humans and animals:
  {Convergence} towards a transformation account of hippocampal--neocortical
  interactions.
\newblock {\em Neuropsychologia}, 48(8):2339--2356.

\bibitem[Xie et~al., 2013]{xie2013sleep}
Xie, L., Kang, H., Xu, Q., Chen, M.~J., Liao, Y., Thiyagarajan, M., O'Donnell,
  J., Christensen, D.~J., Nicholson, C., Iliff, J.~J., et~al. (2013).
\newblock Sleep drives metabolite clearance from the adult brain.
\newblock {\em science}, 342(6156):373--377.

\bibitem[Yamins et~al., 2014]{yamins_performance-optimized_2014}
Yamins, D. L.~K., Hong, H., Cadieu, C.~F., Solomon, E.~A., Seibert, D., and
  DiCarlo, J.~J. (2014).
\newblock Performance-optimized hierarchical models predict neural responses in
  higher visual cortex.
\newblock {\em Proceedings of the National Academy of Sciences},
  111(23):8619--8624.

\bibitem[Yee et~al., 2013]{yee_semantic_2013}
Yee, E., Chrysikou, E.~G., and Thompson-Schill, S.~L. (2013).
\newblock {\em Semantic {Memory}}.
\newblock Oxford University Press.

\bibitem[Zbontar et~al., 2021]{zbontar2021barlow}
Zbontar, J., Jing, L., Misra, I., LeCun, Y., and Deny, S. (2021).
\newblock Barlow twins: Self-supervised learning via redundancy reduction.
\newblock {\em arXiv preprint arXiv:2103.03230}.

\bibitem[Zeiler and Fergus, 2013]{zeiler_visualizing_2013}
Zeiler, M.~D. and Fergus, R. (2013).
\newblock Visualizing and {Understanding} {Convolutional} {Networks}.
\newblock {\em arXiv:1311.2901 [cs]}.
\newblock arXiv: 1311.2901.

\bibitem[Zhuang et~al., 2021]{zhuang_unsupervised_2021}
Zhuang, C., Yan, S., Nayebi, A., Schrimpf, M., Frank, M.~C., DiCarlo, J.~J.,
  and Yamins, D. L.~K. (2021).
\newblock Unsupervised neural network models of the ventral visual stream.
\newblock {\em Proceedings of the National Academy of Sciences},
  118(3):e2014196118.

\end{thebibliography}


\newpage


\section{Supplementary material}

\subsection{Training losses for full and pathological models}

\begin{figure}[t]
  {\centering
    \includegraphics[width=0.9\textwidth]{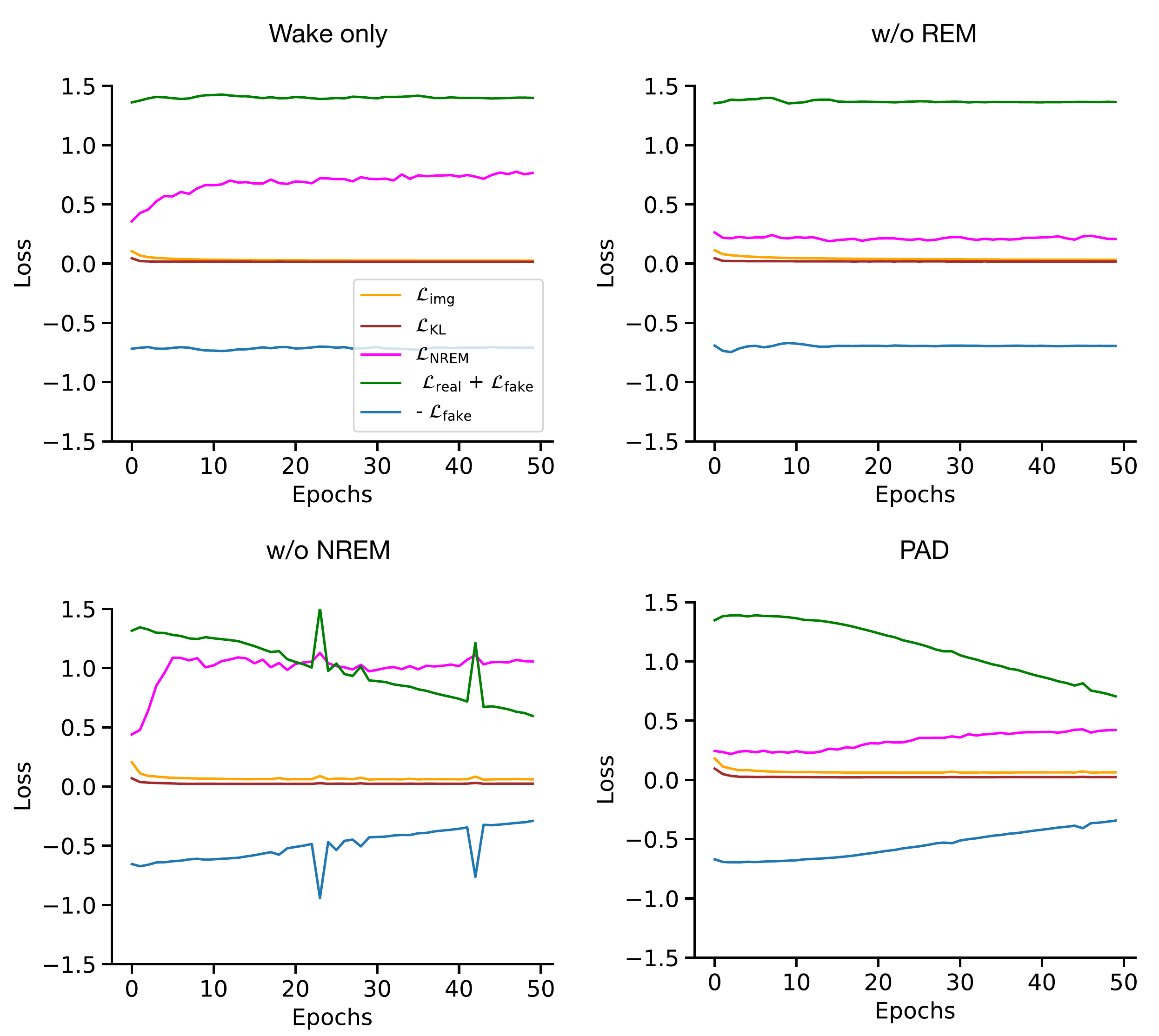}
    \caption{{\bf Training losses for full and pathological models with CIFAR-10 dataset.}
    Evolution of training losses used to optimize $E$ and $G$ networks (see Methods) over training epochs for full and pathological models.
    }\label{fig:supp_losses_cifar10}
    \medskip}
  \small
\end{figure}

\begin{figure}[t]
  {\centering
    \includegraphics[width=0.9\textwidth]{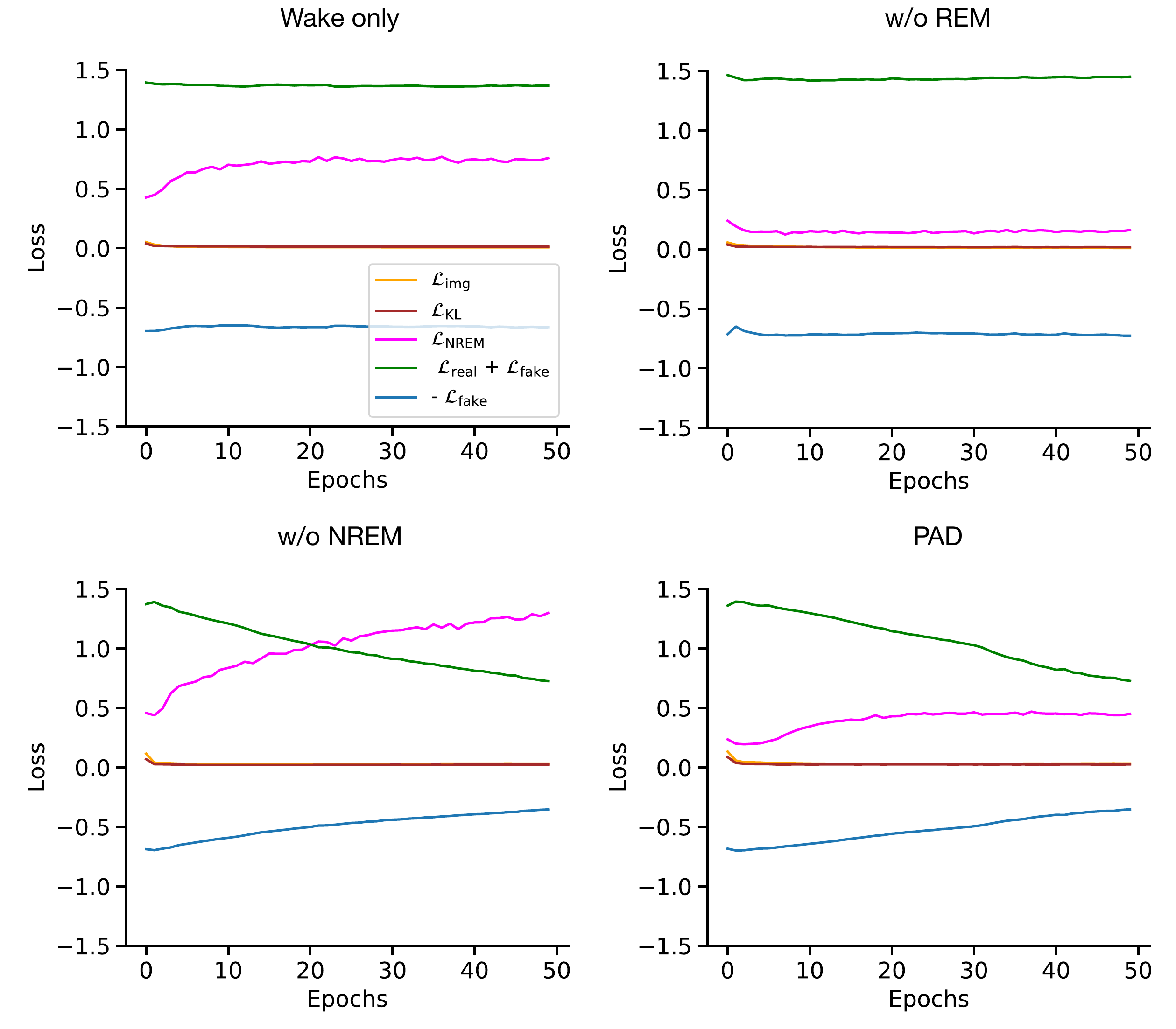}
    \caption{{\bf Training losses for full and pathological models with SVHN dataset.}
    }\label{fig:supp_losses_svhn}
    \medskip}
  \small
\end{figure}

In the following, we report the measured losses over training for the various different pathological conditions.
$\mathcal{L}_{\text{img}}$ and $\mathcal{L}_{\text{KL}}$ are optimized for each condition and systematically decrease with learning, while $\mathcal{L}_{\text{NREM}}$ is significantly reduced in models with NREM (Figs.~\ref{fig:supp_losses_cifar10}, \ref{fig:supp_losses_svhn}).
Its initial increase in the models with REM is explained to its competitive optimization with the GAN losses.
Generator loss $\mathcal{L}_{\text{fake}} = \mathcal{L}_{\text{REM}}$ and discriminator loss $\mathcal{L}_{\text{real}} + \mathcal{L}_{\text{fake}}$ are only optimized in models with REM, showing a progressive decrease of the discriminator loss in parallel with an increase of the generator loss, reflecting adversarial learning between the two streams.

\subsection{Linear classification performance}

We report the mean and standard error of the mean (SEM) of the final linear classification performance (epoch 50) on latent representations of from the PAD and pathological models in Table~\ref{tab:linear}.

\begin{table}[t]
\begin{tabular}{|l | l l l l l|} 
 \hline
Dataset & PAD & w/o memory mix & w/o REM & w/o NREM & Wake only   \\ 
 \hline
CIFAR-10 & $58.25 \pm 0.70$ & $53.87 \pm 0.85$ & $46.00 \pm 0.43$ & $58.00 \pm 0.34$  & $42.25 \pm 0.54$ \\
 \hline
SVHN & $78.92 \pm  0.40$ & $60.87 \pm 5.07$ & $42.30 \pm 1.51$ & $73.25 \pm  0.22$  & $41.93 \pm 0.65$ \\
 \hline
\end{tabular}
\caption{\label{tab:linear} {\bf Final classification performance for full model and all pathological conditions for un-occluded images .}
Mean and SEM over 4 different initial condition of linear separability of latent representations at the end of training (epoch 50) for PAD and its pathological variants.}
\end{table}

\begin{figure}[t]
  {\centering
    \includegraphics[width=0.9\textwidth]{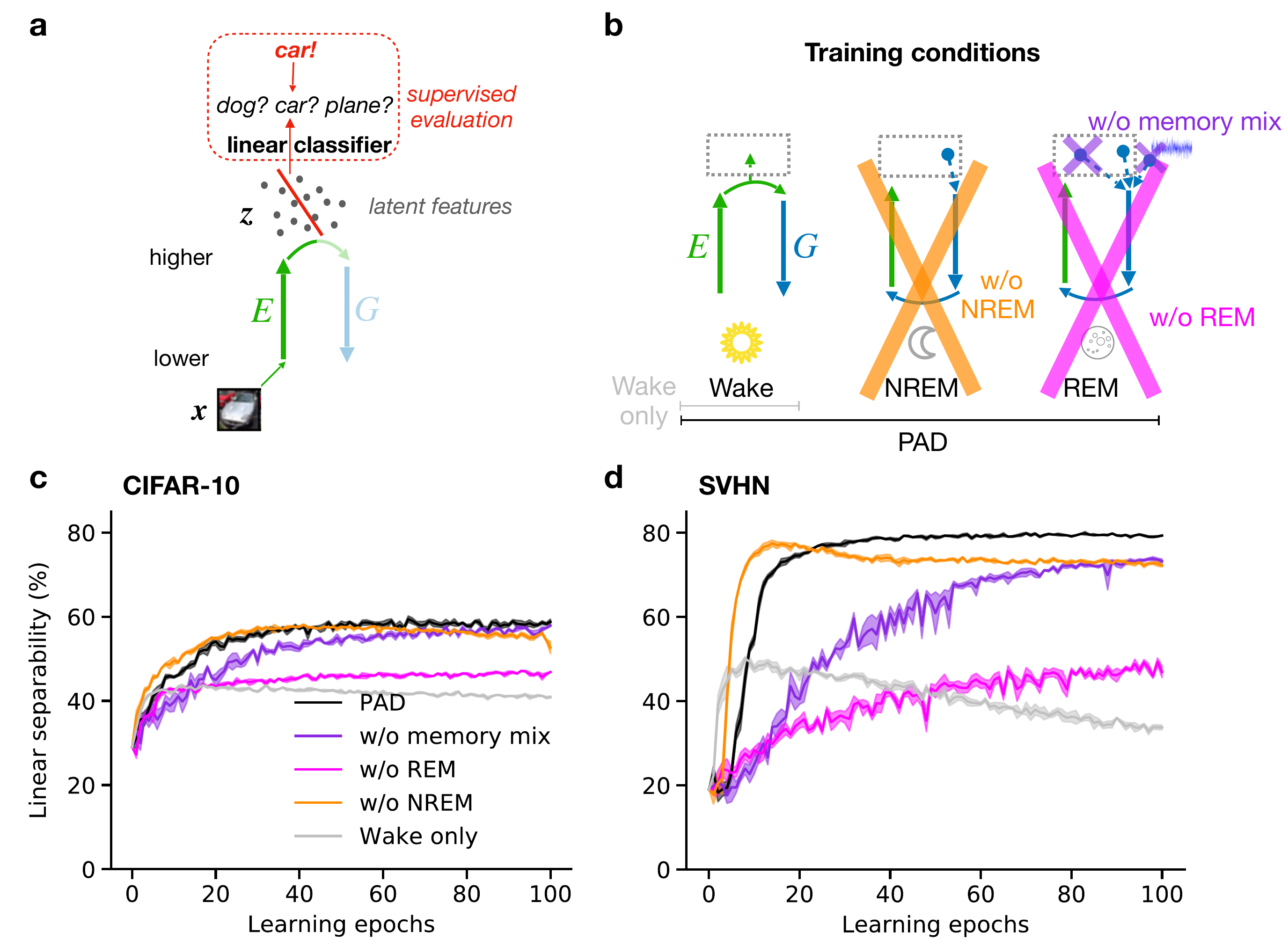}
    \caption{{\bf Linear classification performance for full model and all pathological conditions.}
      For details see Fig.~\ref{fig:results_linear_REM}.
    }
    \label{fig:supp_linear}
    \medskip}
  \small
\end{figure}

We also report the linear classification performance for the full and pathological models over $100$ epochs.
Linear separability for the "w/o REM" (Figs.~\ref{fig:supp_linear}c,d, pink curves) and "w/o memory mix" (Figs.~\ref{fig:supp_linear}d, purple curve) conditions do not reach levels of the full model (Figs.~\ref{fig:supp_linear}c,d, black curves) even after many training epochs.
Furthermore, without NREM (Figs.~\ref{fig:supp_linear}c,d, "w/o NREM" and "Wake only", orange and gray curves), linear separability tends to decrease after many training epochs, suggesting that NREM helps to stabilize performance with training by preventing overfitting.

\subsection{Comparison of performance with REM driven by convex combination or noise}

\begin{figure}[t]
  {\centering
    \includegraphics[width=0.9\textwidth]{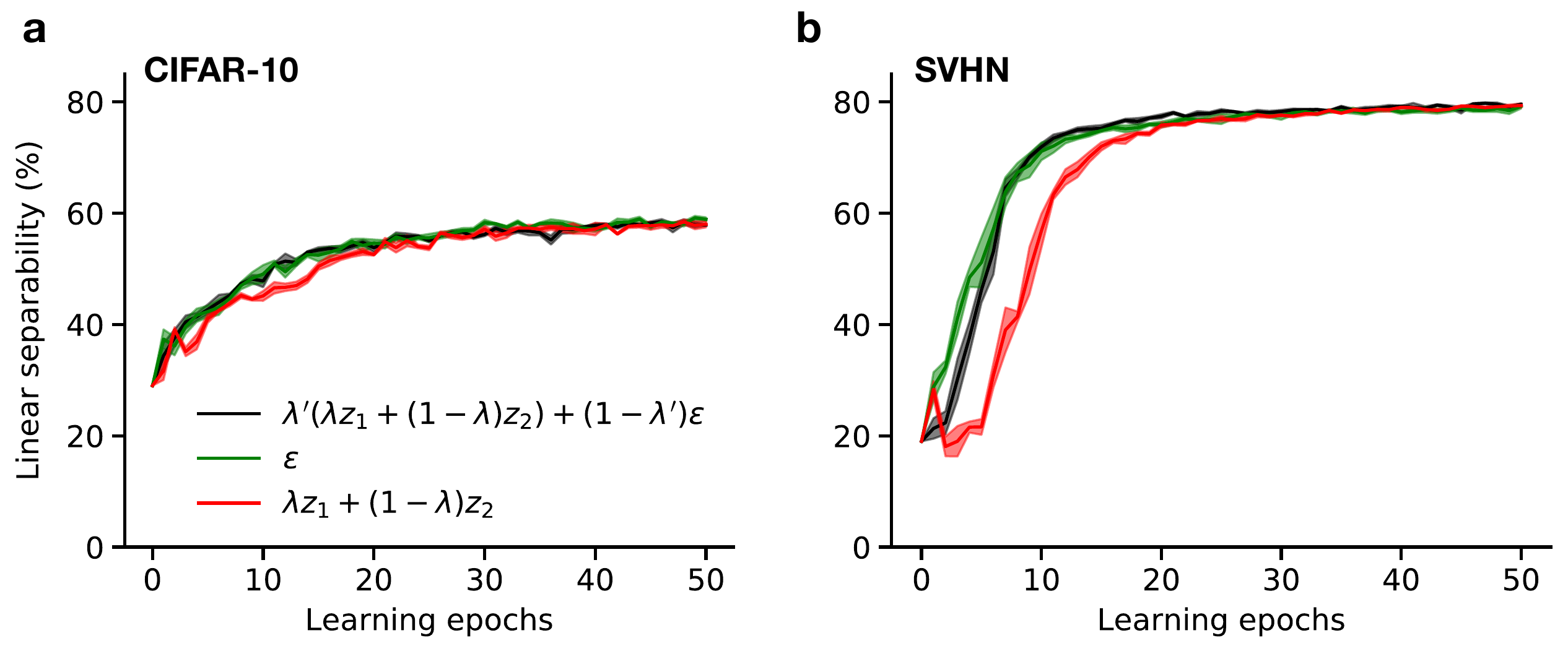}
    \caption{{\bf Linear classification performance for different mixing strategies during REM.}
      Linear separability of latent representations with training epochs for PAD trained with different REM phases: one driven by a convex combination of mixed memories and noise (black), one by pure noise (green), and one by mixed memories only (red). For details see Fig.~\ref{fig:results_linear_REM}. 
    }
    \label{fig:supp_linear_gaussian}
    \medskip}
\end{figure}

We report the linear classifier performance for PAD using different latent inputs to the generator. In the main text, we use a convex combination of mixed memories (being a convex combination of two different replayed latent vectors) and noise sampled from a Gaussian unit distribution (Fig.~\ref{fig:supp_linear_gaussian}, black). We here show the results when only random Gaussian noise is used (Fig.~\ref{fig:supp_linear_gaussian}, green) and when only a convex combination of memories is used (Fig.~\ref{fig:supp_linear_gaussian}, red).
These different mixing strategies do not show a big difference in linear separability over training epochs.

\subsection{The order of sleep phases has no influence on the performance of the linear classifier}

\begin{figure}[t]
  {\centering
    \includegraphics[width=0.9\textwidth]{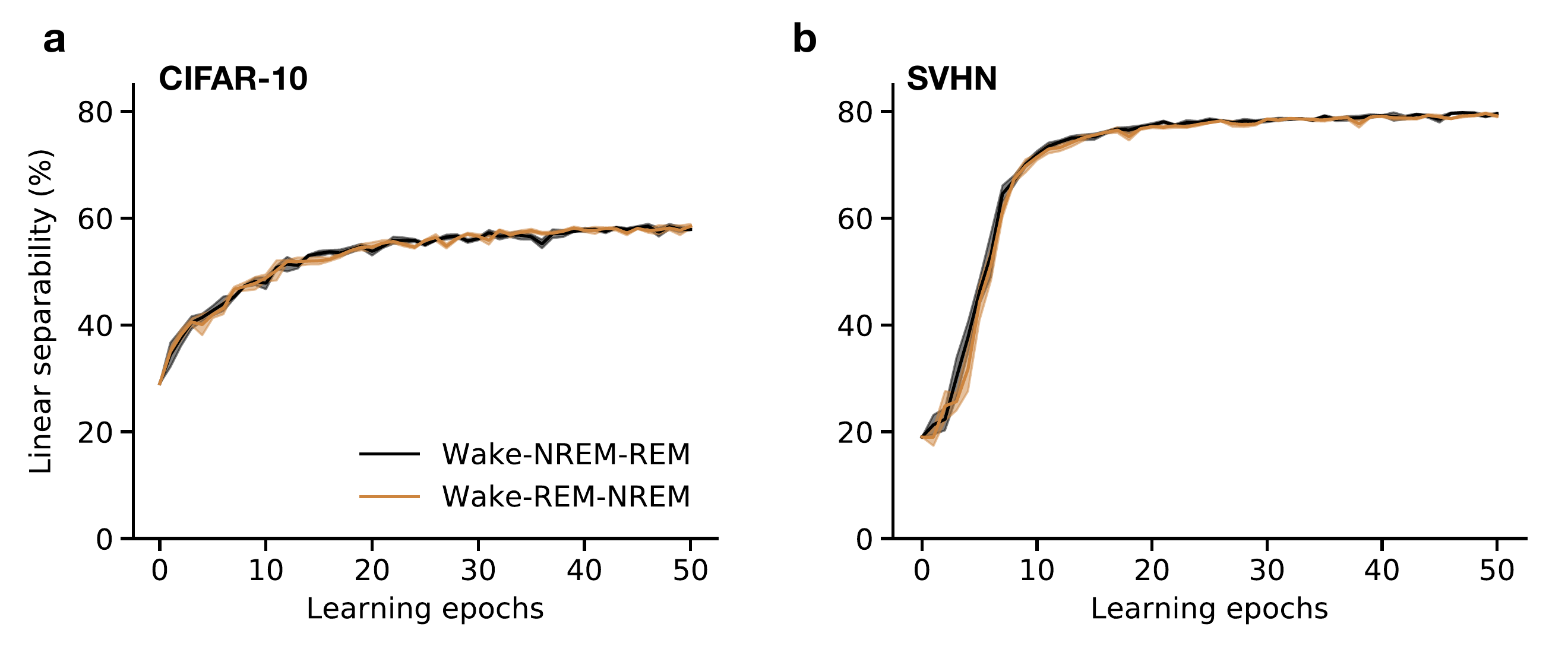}
    \caption{{\bf Linear classification performance for different order of sleep phases.}
      Linear separability of latent representations with training epochs for PAD trained when NREM precedes REM phase (Wake-NREM-REM, black) or when REM precedes NREM (Wake-REM-NREM, brown).
    }
    \label{fig:supp_linear_order}
    \medskip}
\end{figure}

To investigate the role of the order of NREM and REM sleep phases, we consider a variation in which their order is reversed with respect to the model described in the main manuscript.
The performance of the linear classifier is not influenced by this change (Fig.~\ref{fig:supp_linear_order}).

\subsection{Replaying multiple episodic memories during NREM sleep}

\begin{figure}[t]
  {\centering
    \includegraphics[width=0.9\textwidth]{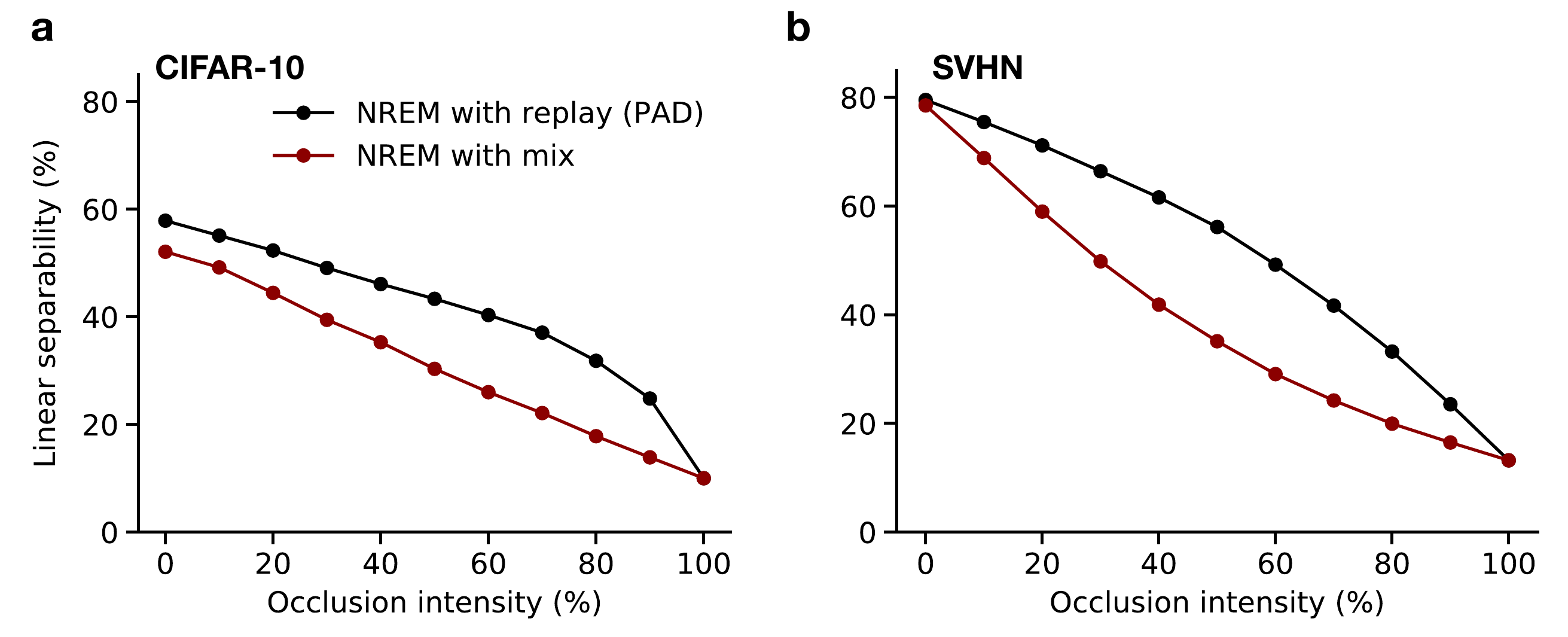}
    \caption{{\bf Importance of replaying single hippocampal memories during NREM.}
    Linear separability of latent representations at the end of learning with occlusion intensity for a model trained with all phases.
    }\label{fig:supp_nrem_mix}
    \medskip}
  \small
\end{figure}

While in the main text we considered NREM to use only a single episodic memory, here we report results for a model in which also NREM uses multiple (here: two) episodic memories.
In the full model (Fig.~\ref{fig:supp_nrem_mix}, black curves, same data as in Fig.~\ref{fig:results_linear_NREM}c,d), NREM uses a single stored latent representation. Here we additionally consider an additional model in which these representations are obtained from a convex combination of mixed memories and spontaneous cortical activity.
The better performance of a single replay suggests that replay from single episodic memories as postulated to occur during NREM sleep is more efficient to robustify latent representations against input perturbations.

\end{document}